\documentclass[reprint,aps,prd,superscriptaddress,preprintnumbers,nofootinbib]{revtex4-1}

\usepackage{amsfonts}
\usepackage{amsmath}
\usepackage{amsthm}
\usepackage{amssymb}
\usepackage{array}
\usepackage{dcolumn}
\usepackage[svgnames]{xcolor}
\usepackage[hidelinks]{hyperref}
\hypersetup{
    colorlinks=true,
    linkcolor=NavyBlue,
    filecolor=NavyBlue,
    urlcolor=NavyBlue,
    citecolor=NavyBlue,
    }

\usepackage{graphics}
\usepackage{tikz}
\usepackage{graphicx}
\usetikzlibrary{positioning}
\usepackage{caption}
\usepackage{subcaption}
\usepackage{adjustbox}
\usepackage{gensymb}
\usepackage{breqn}
\usepackage{braket}
\usepackage{enumitem}
\usepackage{gensymb}

\begin{document}

\hfill \preprint{ MI-TH-2012}

\title{Explaining $(g-2)_{\mu,e}$, the KOTO anomaly and the MiniBooNE excess in an extended Higgs model with sterile neutrinos}

\author{Bhaskar Dutta}
\email{dutta@physics.tamu.edu}
\author{Sumit Ghosh}
\email{ghosh@tamu.edu}
\affiliation{Mitchell Institute for Fundamental Physics and Astronomy$,$ Department~ of ~Physics ~ and~ Astronomy$,$\\ Texas A$\&$M University$,$~College~ Station$,$ ~Texas ~77843$,$~ USA}

\author{Tianjun Li}
\email{tli@itp.ac.cn}
\affiliation{CAS Key Laboratory of Theoretical Physics$,$~ Institute~ of~ Theoretical ~Physics$,$ \\ Chinese Academy of Sciences$,$~ Beijing$,$~ 100190$,$~ People's~ Republic ~China}
\affiliation{School of Physical Sciences$,$~ University~ of ~Chinese~ Academy~ of ~Sciences$,$  \\ Beijing 100049$,$~ People's ~Republic ~China}

\begin{abstract}
 We consider a simple extension of the Standard Model (SM) by a complex scalar doublet and a singlet along with three sterile neutrinos.  The sterile neutrinos mix with the SM neutrinos to produce three light neutrino states consistent with the oscillation data and three  heavy sterile states. The lightest sterile neutrino has lifetime longer than the age of the Universe and can provide correct dark matter relic abundance. Utilizing  tree-level flavor changing interactions of a light scalar with mass~$\sim\mathcal{O}(100)$~MeV along with  sterile neutrinos, we can explain the anomalous magnetic moments of both muon and electron, KOTO anomalous events and the MiniBooNE excess simultaneously.  
\end{abstract}

\maketitle


\section{Introduction} \label{sec:introduction}
The Standard Model~(SM) of particle physics is a very successful, mathematically consistent theory of the known elementary particles. Most of the SM predictions are consistent with the experimental data. However, some theoretical puzzles and experimental results cannot be explained solely based on the SM. These  are the hints that we need some new physics beyond the SM. The need for new physics beyond the SM is well established in the neutrino sector of the SM where the   neutrino oscillation data~\cite{Fukuda:1998mi,Ahmad:2002jz} definitely
require at least two neutrinos to have non-zero masses. On the other hand, the SM does not provide any dark matter (DM) candidate which could explain the observed DM content of the Universe~\cite{Aghanim:2018eyx}. In addition to the neutrino and DM puzzles,  a few other experimental results associated with the quarks and charged leptons also pose challenges to the SM. 

 The anomalous magnetic moment of the muon is one of the long-standing deviations of
the experimental data from the theoretical predictions of the SM. There exists  a 3.7 $\sigma$  discrepancy between the experimental results~\cite{Bennett:2006fi, Tanabashi:2018oca} and theoretical predictions~\cite{Davier:2017zfy, Blum:2018mom, Keshavarzi:2018mgv, Davier:2019can}.
  This was recently accompanied by a 2.4 $\sigma$ discrepancy between the experimental~\cite{Hanneke:2010au, Hanneke:2008tm} and theoretical~\cite{Aoyama:2017uqe} values of the anomalous magnetic moment of the electron 
due to a recent precise measurement of the fine structure constant~\cite{Parker:2018vye}. It is interesting to note that the deviations  are in opposite directions, and  $\Delta a_e/\Delta a_\mu$ does not follow the lepton mass scaling $m_e^2/m_\mu^2\sim 2.25\times 10^{-5}$. It would require   a model with new flavor structure in the leptonic sector  to explain these discrepancies. Universal flavor structure requires very large Yukawa coupling to explain the anomalies~\cite{Hiller:2019mou}. More data is needed to confirm the $\Delta a_e$ discrepancy. There will be new results for the $a_\mu$ measurement from the Fermilab soon. Very recently, the lattice calculation for  the hadronic light-by-light scattering contribution confirms the $\Delta a_\mu$ discrepancy~\cite{Blum:2019ugy}. Recently, the measurement of the radiative corrections to the pion form factor also confirm the need of a beyond SM explanation of $a_\mu$~\cite{Campanario:2019mjh}.

Any observations of the flavor changing rare decays of kaons also indicate  new physics beyond the SM. One very interesting development in this topic is the recent results from the KOTO experiment which is indicating that $K_L\rightarrow\pi^0\bar{\nu}\nu$ decay takes place at a higher rate compared to the SM prediction~\cite{Shinohara, Lin}. 
The  branching ratio is estimated to be at least two orders of magnitude larger than the SM prediction~\cite{Buras:2015qea}. Any new physics explanation of this excess is, however, constrained by the charged kaon  decay mode $K^+\rightarrow\pi^+\nu \bar{\nu}$ and $K^+\rightarrow\pi^+ X$ which are being investigated at   NA62~\cite{Ruggiero} and E949~\cite{Artamonov:2009sz} experiments, respectively. The new physics to explain the anomaly also requires flavor violating interactions in the quark sector.

The interesting question is can any simple extension of the SM  explain all these observations? In an attempt to find the answer to this question, we propose a simple extension of the SM which contains an additional scalar doublet, a singlet, and three sterile neutrinos. This Higgs sector extension is simple, well-motivated, and is associated with the electroweak sector of the SM~\cite{Branco:2011iw}. We investigate the most general renormalized scalar potential utilizing the electroweak symmetry breaking and explore the parameter space associated with the masses and mixings of the Higgs bosons. The interesting feature of this parameter space is the emergence of a  light scalar that has tree-level flavor violating couplings to the SM fermions.  Further, the sterile neutrinos would help us to realize tiny neutrino masses utilizing type I seesaw in this model. The lightest sterile neutrino can be a viable DM candidate. Utilizing the flavor violation  in the lepton sector, we explain the $g-2$ of both muon and electron. The quark sector flavor violation leads to tree level decays of kaon into pion and dark matter pair which will mimic the $K_L\rightarrow\pi^0\bar{\nu}\nu$ decay channel inside the KOTO detector and help to explain the KOTO anomaly.

In addition to the light neutrino masses and KOTO anomaly, the existence of the sterile neutrinos would help us to explain two other puzzles. One of them is the DM content of the Universe  which can be  explained by the DM candidate in this model, i.e., the lightest sterile neutrino. The other one is the recent MiniBooNE observation where the data exhibits  a 4.8 $\sigma$ excess~\cite{Aguilar-Arevalo:2018gpe, Aguilar-Arevalo:2020nvw} of events over the known background. This excess can be explained  with the muon neutrino getting upscattered to a heavy sterile neutrino  due to the light scalar.

Finally, the parameter space of this light scalar with couplings to leptons and quarks is constrained by various proton, electron beam dump, and collider experiments, lepton flavor violating decays, kaon mixing, and astrophysical data. We explore various constraints and  determine the allowed parameter space where all the anomalies can be explained simultaneously. We also make predictions of this allowed parameter space  for various ongoing and upcoming experiments.

The rest of the paper is organized as follows:  In Sec.~\ref{sec:model} we discuss the model by defining necessary   parameters and  interaction terms. The origin of  neutrino mass is presented in Sec.~\ref{sec:neutrino mass}. In Sec.~\ref{sec:dark matter}, we discuss the possibility of the lightest sterile neutrino as a DM candidate. We generate a viable physical scalar spectrum in Sec.~\ref{sec:light scalar}.   In Sec.~\ref{sec:g-2}, we study the anomalous magnetic moments of the electron and muon and allowed parameter space. In Sec.~\ref{sec:koto}, we discuss the allowed parameter space associated with the KOTO anomaly. In Sec.~\ref{sec:miniboone} V, we discuss the recent MiniBooNE observation. We summarize our analysis in Sec.~\ref{sec:discussions} by showing a few benchmark points (BP) which explain all the anomalies after satisfying all other experimental data. We provide additional pieces of information in the Appendices.

\section{model} \label{sec:model}

The scalar sector of the SM has the simplest possible structure with one scalar doublet~\cite{Higgs:1964pj, Higgs:1966ev, Englert:1964et, Guralnik:1964eu, Kibble:1967sv}.    Two-Higgs-doublet model (2HDM)~\cite{Lee:1973iz, Branco:2011iw} and its singlet/triplet extensions are well motivated  extension of the SM scalar sector~\cite{He:2008qm, Grzadkowski:2009iz, Logan:2010nw, Boucenna:2011hy, He:2011gc, Bai:2012nv, He:2013suk, Cai:2013zga, Guo:2014bha, Wang:2014elb, Drozd:2014yla, Campbell:2015fra, Drozd:2015gda, vonBuddenbrock:2016rmr, Muhlleitner:2016mzt, Liu:2015oaa}. In this work, we consider a simple extension of the CP-conserving 2HDM by adding one complex scalar singlet.  In addition to this, we extend the SM fermion sector by adding three right-handed sterile neutrinos $n^\prime_{R_i}$ with $i=1,2,3$ to explain the observed neutrino masses and mixings.  The quantum numbers of the scalars  under the SM gauge group $SU(2)_L\times U(1)_Y$ are 
\begin{equation}
\phi_1\sim (2,1/2),~~~~~\phi_2\sim (2,1/2),~~~~~\phi_S\sim (1,0)~,~\,
\end{equation} \\
and the definition of the electric charge is $Q\equiv T_3+Y$.

In general, the scalar sector can be CP-violating. For simplicity, we assume that the scalar sector respects the CP symmetry. Also, we do not impose any discrete symmetry. The most general renormalizable and CP-conserving scalar potential can be written as follows 

\small{\begin{dmath} \label{potential}
V = m_1^2\phi_1^\dagger \phi_1+ m_2^2\phi_2^\dagger \phi_2+m_{12}^2(\phi_1^\dagger \phi_2+\phi_2^\dagger \phi_1)+ m_S^2\phi_S^\dagger \phi_S-m_{S^\prime}^2 (\phi_S^2+{\phi^{\dagger 2}_S})+m_{1S}(\phi_1^\dagger\phi_1\phi_S+\phi_1^\dagger\phi_1\phi_S^\dagger)+m_{2S}(\phi_2^\dagger\phi_2\phi_S+\phi_2^\dagger\phi_2\phi_S^\dagger)+ \frac{\lambda_1}{2}(\phi_1^\dagger \phi_1)^2+\frac{\lambda_2}{2}(\phi_2^\dagger \phi_2)^2+
\frac{\lambda_S}{2}(\phi_S^\dagger \phi_S)^2+\lambda_3 (\phi_1^\dagger \phi_1)( \phi_2^\dagger \phi_2)+\lambda_4 (\phi_1^\dagger \phi_2)( \phi_2^\dagger \phi_1)+\lambda_5\left[(\phi_1^\dagger \phi_2)^2 +(\phi_2^\dagger \phi_1)^2 \right]+\lambda_6 \left[ (\phi_1^\dagger \phi_1)(\phi_1^\dagger \phi_2)+(\phi_1^\dagger \phi_1)(\phi_2^\dagger \phi_1) \right]+\lambda_7 \left[ (\phi_2^\dagger \phi_2)(\phi_1^\dagger \phi_2)+(\phi_2^\dagger \phi_2)(\phi_2^\dagger \phi_1) \right]+\lambda_{1S} (\phi_1^\dagger \phi_1 )(\phi_S^\dagger \phi_S)+\lambda_{2S}( \phi_2^\dagger \phi_2)( \phi_S^\dagger \phi_S)+\lambda_{12S}\left[ (\phi_1^\dagger \phi_2)(\phi_S^\dagger \phi_S) + (\phi_2^\dagger \phi_1)(\phi_S^\dagger \phi_S) \right]+m_{12S}( \phi_1^\dagger \phi_2  \phi_S +\phi_S^\dagger \phi_2^\dagger  \phi_1 ) ~.~\, \end{dmath}}

 We choose to work in the Higgs basis~\cite{Georgi:1978ri, Botella:1994cs, Lavoura:1994fv, Donoghue:1978cj, Lavoura:1994yu}, where only one of the doublet $\phi_1$ gets a vacuum expectation value (vev), $\braket{\phi_1}=v/\sqrt{2}$.   The details about the Higgs basis for the scalar structure of our model is given in Appendix~\ref{appendix:higgs basis}. The doublet $\phi_1$ completely controls  the spontaneous electroweak gauge symmetry breaking and the mass generations of the fermions and gauge bosons. While the other doublet and the singlet are ordinary scalars.  In the following, we analyze the scalar sector in the Higgs basis. After the spontaneous symmetry breaking, we can write the scalars as
\begin{eqnarray}
	 &\phi_1 &\sim   \left( \begin{array}{c}  {G}^+  \\  \frac{1}{\sqrt{2}}(v +\rho_1+iG_0)  \end{array} \right), \phi_2 \sim   \left( \begin{array}{c}  {\phi_2}^+ \nonumber \\  \frac{1}{\sqrt{2}}(\rho_2+i\eta_2)  \end{array} \right),~~~~\\&\phi_S&\sim \frac{1}{\sqrt{2}}(\rho_S+i\eta_S)~.~\,
\end{eqnarray}
 The extremization of the potential in Eq.~\ref{potential} gives the following conditions \begin{eqnarray}
 m_1^2+\frac{\lambda_1 v^2}{2}&=&0 ~,~\,\\ \label{tadpole} m_{12}^2+\frac{\lambda_6 v^2}{2}&=&0~.~\,
 \end{eqnarray}   
 Eq.~\ref{tadpole} makes sure that the $\phi_2$ does not get a vev. From the minimizing  conditions, we further  get \begin{eqnarray} &\lambda_1 &>0,~~~~m_1^2<0, ~~~\lambda_5>0,~~~\lambda_6>0,~~~m_{12}^2<0, \nonumber \\ &m_{12S}&>0,~~~m_{1S}=0~.~\, \end{eqnarray} 
 The vev of $\phi_S$ is zero due to $m_{1S}=0$. Therefore, the total number of free parameters in the scalar sectors is 17 including the vev $v$. The total number of scalar degrees of freedom (dof) is 10. Three dof get eaten to give mass to $W^{\pm} ~\mbox{and}~ Z$ gauge bosons. The remaining 7 are physical Higgs. In the Higgs basis, $G^\pm$ and $G_0$ become the Goldstone bosons. $\phi_2^\pm$ gives two charged physical Higgs $h^\pm$. CP-even states $\rho_1$,  $\rho_2$ and $\rho_S$ mix to give three neutral physical scalars $h$, $h_1$ and $h_2$. We identify the $h$ as the SM Higgs boson. The CP-odd states $\eta_2$ and $\eta_S$ mix and gives two neutral physical pseudoscalar $s_1$ and $s_2$. 

The physical charged scalar mass is given by  \begin{equation} \label{chargedscalar} m_{h^\pm}^2 = m_2^2+\frac{\lambda_3v^2}{2}~.~\,\end{equation}

The mixing of the three CP-even neutral scalars $\rho_1$, $\rho_2$ and $\rho_S$ is  \small{\begin{equation} V^\rho_{mass}=\frac{1}{2}\left(\rho_1 ~~~ \rho_2 ~~~\rho_S  \right)~ \left(M^2_\rho \right)_{3\times3}~ \left( \begin{array}{c} \rho_1 \\ \rho_2  \\ \rho_S\end{array} \right) ~,~\,
 \end{equation} } where the $3 \times 3$ mass square matrix $M^2_\rho$ is 
 \small{\begin{equation}M^2_\rho = \left( \begin{array}{ccc}\lambda_1 v^2&\lambda_6 v^2&0 \\ \lambda_6 v^2& m_2^2+\frac{\lambda^+_{345}v^2}{2}&\frac{m_{12S}v}{\sqrt{2}} \\ 0&\frac{m_{12S}v}{\sqrt{2}} &m_S^2-2 m_{S^\prime}^2+\frac{\lambda_8v^2}{2} \end{array} \right) ~.~\, \end{equation}}
Here, we have used Eq.~\ref{tadpole} to simplify terms in the mass squared matrix and defined $\lambda^+_{345} \equiv \lambda_3+\lambda_4+\lambda_5 $. We get three physical scalars from this mixing, $h$, $h_1$ and $h_2$ with mass squared $m_h^2, m_{h_1}^2$ and $m_{h_2}^2$, respectively. The fields in the mass basis, $h, h_1$ and $h_2$ are related to those in the interaction basis, $\rho_1,\rho_2$ and $\rho_S$ by a $3\times 3$ rotation matrix ${U_{R}}_{3\times3}(\theta_i)$ which can be parametrized with three Euler angles $\theta_1,\theta_2 ~\mbox{and} ~\theta_3$. We write $U_{R}$  as follows
\small{\begin{eqnarray} \label{3x3mixing} U_{R} &=& \left( \begin{array}{ccc} c_{11} & c_{12} & c_{13} \\ c_{21} & c_{22} & c_{23} \\ c_{31} & c_{32} & c_{33} \end{array} \right) ~,~\,\end{eqnarray}}
where $\rho_i = {U_{R}}_{ij} h_j $.  The quantities $c_{ij}$ are functions of  $\cos \theta_k$ and $\sin \theta_k$ ($k=1,2,3$).  The interaction states can be written in terms of the physical states as 
\begin{eqnarray} \rho_1 &=& c_{11} h_2 + c_{12} h + c_{13} h_1 ~,~\,\nonumber\\
\rho_2 &=&  c_{21} h_2 + c_{22} h + c_{23} h_1 ~,~\,\nonumber\\ 
\rho_S &=&  c_{31} h_2 + c_{32} h + c_{33} h_1~.~\,\end{eqnarray}
 
 The mixing of the two CP-odd neutral scalars $\eta_2$-$\eta_S$ can be written as
 \small{\begin{eqnarray} V^\eta_{mass}=\frac{1}{2}\left( \eta_2 ~~~\eta_S  \right) \left( M^2_\eta \right)_{2\times 2}\left( \begin{array}{c} \eta_2 \\ \eta_S \end{array}  \right)~,~\,
 \end{eqnarray}} 
 where the $2 \times 2$ mass square matrix $M^2_\eta$ is given by
 \small{\begin{equation}M^2_\eta =  \left( \begin{array}{cc} m_2^2+\frac{\lambda^-_{345}v^2}{2}& -\frac{m_{12S}v}{\sqrt{2}} \\ -\frac{m_{12S}v}{\sqrt{2}} &m_S^2+2 m_{S^\prime}^2+\frac{\lambda_8v^2}{2} \end{array} \right) ~,~\,\end{equation}} where we define $\lambda^-_{345} \equiv \lambda_3+\lambda_4-\lambda_5 $. From the above mixing, we get two physical neutral pseudoscalar
 \begin{equation} \left( \begin{array}{c} s_1 \\ s_2 \end{array} \right) =  \left( \begin{array}{cc} \cos \alpha & -\sin  \alpha \\ \sin  \alpha &\cos  \alpha \end{array} \right) \left( \begin{array}{c} \eta_2 \\ \eta_S \end{array} \right) ~,~\,\end{equation} where the mixing angle is given by  \begin{equation} \tan 2 \alpha = \frac{m_{12S}v/\sqrt{2}}{m_{11}^{ 2}-m_{22}^2} \end{equation} with the corresponding mass squared
\small{\begin{equation}  m_{s_1}^2=\frac{1}{2}(m^{2}_{11}+m^2_{22})-\frac{1}{2}\sqrt{(m^{2}_{11}-m^2_{22})^2+\frac{m_{12S}^2v^2}{2}}  \end{equation}} and  \small{\begin{equation}  m_{s_2}^2=\frac{1}{2}(m^{2}_{11}+m^2_{22})+\frac{1}{2}\sqrt{(m^{2}_{11}-m^2_{22})^2+\frac{m_{12S}^2v^2}{2}}  ~,~\,
\end{equation}} respectively, where \small{\begin{equation} m^{2}_{11} = \frac{1}{2} \left( m_2^2+\frac{\lambda_3v^2}{2}+\frac{\lambda_4v^2}{2}-\frac{\lambda_5v^2}{2} \right)\end{equation}} and \small{\begin{equation} m^2_{22} = \frac{1}{2} \left(m_S^2+2 m_{S^\prime}^2+\frac{\lambda_8v^2}{2} \right)~.~\,  \end{equation}}
 The interaction states can be written in terms of the mass eigenstates as
 \begin{eqnarray} \label{etadef} \eta_2 &=& \cos\alpha~ s_1 +\sin\alpha~s_2 ~,~\,\nonumber\\ \eta_S &=& -\sin\alpha ~s_1 +\cos\alpha~s_2  ~.~\,\end{eqnarray}

Both  scalar doublets interact with all the fermions in the interaction basis, while the singlet scalar only interacts with the sterile neutrinos. The masses of the fermions come from the interactions with $\phi_1$. The couplings of $\phi_2$ to the fermions are unconstrained and do not need to respect the SM fermion flavor symmetry. Therefore, the interactions of the fermions with the neutral components of $\phi_2$ can generate the tree-level flavor-changing neutral current (FCNC), which would be useful to explain the KOTO anomaly and g-2 of the electron. The fermions can interact with the singlet scalar through the scalar mixings discussed above. The complete Yukawa sector Lagrangian in the interaction basis is 
\small{ \begin{eqnarray} \label{yukawa:int}
 -\mathcal{L} &=&  \bar{q}^\prime_{L_i}(y^\prime_{1d})_{ij} d^\prime_{R_j} \phi_1 +\bar{q}^\prime_{L_i}(y^\prime_{1u})_{ij} u^\prime_{R_j} \tilde{\phi_1}\nonumber\\* &+&\bar{l}^\prime_{L_i}(y^\prime_{1e})_{ij} e^\prime_{R_j} \phi_1 +\bar{l}^\prime_{L_i}(y^\prime_{1n})_{ij} n^\prime_{R_j} \tilde{\phi_1} \nonumber\\* &+& \bar{q}^\prime_{L_i}(y^\prime_{2d})_{ij} d^\prime_{R_j} \phi_2 +\bar{q}^\prime_{L_i}(y^\prime_{2u})_{ij} u^\prime_{R_j} \tilde{\phi_2}\nonumber\\* &+&\bar{l}^\prime_{L_i}(y^\prime_{2e})_{ij} e^\prime_{R_j} \phi_2 + \bar{l}^\prime_{L_i}(y^\prime_{2n})_{ij} n^\prime_{R_j} \tilde{\phi_2} \nonumber\\* &+& \bar{n}^{\prime c}_{R_i} (y^\prime_{sn})_{ij}n^\prime_{R_j}\phi_S + \frac{1}{2}\bar{n}^{\prime c}_{R_i} M^\prime_{ij}n^\prime_{R_j}+H.c.~,~\, \end{eqnarray}}
   where $i,j$ are the family indices,  $i,j=1,2,3$, and $a,b=1,2$. The primed fermions are the fermions in the interaction basis. The first four terms give the down-type quark masses, up-type quark masses, charged lepton masses, and Dirac mass terms of neutrino, respectively.  The last term gives the Majorana mass terms for the right-handed neutrinos.  In general, all the Yukawa couplings are $3 \times 3$ complex matrices. 
 
 In general, the $3\times3$ Yukawa matrices $y^\prime_{1d}$, $y^\prime_{1u}$, $y^\prime_{1e}$ and $y^\prime_{1n}$, and the mass matrix $M^\prime_{ij}$ can be diagonalized through biunitary transformations as follows
 \small{ \begin{eqnarray} \label{def:diag} & &U^\dagger_{d_{L}} y^\prime_{1d} U_{d_{R}} = y_{1d},~~~~~\mbox{with}~~~(y_{1d})_{ij} = (y_{1d})_{ii} \delta_{ij} ~,~\,\\
  & &U^\dagger_{u_{L}} y^\prime_{1u} U_{u_{R}} = y_{1u},~~~~~\mbox{with}~~~(y_{1u})_{ij} = (y_{1u})_{ii} \delta_{ij} ~,~\,\\
  & &U^\dagger_{e_{L}} y^\prime_{1e} U_{e_{R}} = y_{1e},~~~~~\mbox{with}~~~(y_{1e})_{ij} = (y_{1e})_{ii} \delta_{ij} ~,~\,\\
  & &U^\dagger_{\nu_{L}} y^\prime_{1n} U_{n_{R}} = y_{1n},~~~~~\mbox{with}~~~(y_{1n})_{ij} = (y_{1n})_{ii} \delta_{ij} ~,~\,\\
  & & U^\dagger_{n_{R}}M^\prime U_{n_{R}} = M,~~~~~\mbox{with}~~~M_{ij}=M_{ii} \delta_{ij}~,~\,
\end{eqnarray}  } where $U_{d_{L}}$, $U_{d_{R}}$, $U_{u_{L}}$, $U_{u_{R}}$, $U_{e_{L}}$, $U_{e_{R}}$, $U_{\nu_{L}}$ and $U_{n_{R}}$ are eight appropriate $3\times3$ unitary matrices. These matrices can be used to define the physical states of the fermions,
 \small{ \begin{eqnarray}\label{def:physical}& & d_{L_i} = (U^\dagger_{d_{L}})_{ij}{d}^\prime_{L_j},~~~~~~~~~~~~~ d_{R_i} = (U^\dagger_{d_{R}})_{ij}{d}^\prime_{R_j}~,~\,\\
  & & u_{L_i} = (U^\dagger_{u_{L}})_{ij}{u}^\prime_{L_j},~~~~~~~~~~~~~ u_{R_i} = (U^\dagger_{u_{R}})_{ij}{u}^\prime_{R_j} ~,~\,\\
& & e_{L_i} = (U^\dagger_{e_{L}})_{ij}{e}^\prime_{L_j},~~~~~~~~~~~~~ e_{R_i} = (U^\dagger_{e_{R}})_{ij}{e}^\prime_{R_j} ~,~\,\\
& & \nu_{L_i} = (U^\dagger_{\nu_{L}})_{ij}{\nu}^\prime_{L_j},~~~~~~~~~~~~~ n_{R_i} = (U^\dagger_{n_{R}})_{ij}{n}^\prime_{R_j}~.~\,
\end{eqnarray}}

We also define the following matrices, 
\small{\begin{eqnarray}\label{def:y2}(y_{2d})_{ij} &=& (U^\dagger_{d_{L}})_{ik}(y^\prime_{2d})_{kl}(U_{d_{R}})_{lj}~,~\,\\
(y_{2u})_{ij} &=& (U^\dagger_{u_{L}})_{ik}(y^\prime_{2u})_{kl}(U_{u_{R}})_{lj}~,~\,\\
(y_{2e})_{ij} &=& (U^\dagger_{e_{L}})_{ik}(y^\prime_{2e})_{kl}(U_{e_{R}})_{lj}~,~\,\\
(y_{2n})_{ij} &=& (U^\dagger_{\nu_{L}})_{ik}(y^\prime_{2n})_{kl}(U_{n_{R}})_{lj}~,~\,\\
\label{def:y24} (y_{sn})_{ij} &=& (U^\dagger_{n_{R}})_{ik}(y^\prime_{sn})_{kl}(U_{n_{R}})_{lj}~.~\,
\end{eqnarray}}

 Using the definitions Eq.~\ref{def:diag}-\ref{def:y24} and 
the physical scalar states, the Eq.~\ref{yukawa:int} can be written compactly  as follows
\small{\begin{eqnarray} \label{physicalyukawa}
-\mathcal{L}&=&(m_f)_i \bar{f}_i f_i +(m_{\nu_{d}})_i (\bar{\nu}_{L_i}n_{R_i}+\bar{n}_{R_i}\nu_{L_i}) \nonumber\\* &+& \frac{1}{2} M_i (\bar{n}^c_{R_i}n_{R_i} + \bar{n}_{R_i} n^c_{R_i}) \nonumber \\*
&+&\bar{\nu}_{L_i}(U^\dagger_{PMNS})_{ik} (y_{2e})_{kj}e_{R_j}h^+  \nonumber\\&+&\bar{e}_{R_i}(y_{2e})_{ik}(U_{PMNS})_{kj}\nu_{L_j} h^-\nonumber\\*
&-&\bar{e}_{L_i}(U_{PMNS})_{ik} (y_{2n})_{kj}n_{R_j}h^- \nonumber\\* &-&\bar{n}_{R_i}(y_{2n})_{ik}(U^\dagger_{PMNS})_{kj}e_{L_j} h^+\nonumber\\*
&+&\bar{u}_i[(U_{CKM})_{ik}(y_{2d})_{kj}P_R -(y_{2u})_{ik}(U_{CKM})_{kj}P_L]d_j h^+ \nonumber\\*
&+&\bar{d}_i[(y_{2d})_{ik}(U^\dagger_{CKM})_{kj}P_L-(U^\dagger_{CKM})_{ik}(y_{2u})_{kj}P_R]u_j h^-	\nonumber\\*
&+&\bar{f}_i(y_{f\phi})_{ij} f_j \phi + (y_{n\phi})_{ij} (\bar{\nu}_{L_i} n_{R_j}+\bar{n}_{R_i} \nu_{L_j})\phi \nonumber\\*
&+& (y_{nn\phi})_{ij} (\bar{n}^{ c}_{R_i} n_{R_j}+\bar{n}_{R_i} n^c_{R_j})\phi ~,~\,
\end{eqnarray}}  where $f=d,u,e$; $\phi=h,h_1,h_2,s_1,s_2$ and $(m_f)_i=(y_{1f})_i v / \sqrt{2}$. The Dirac mass matrix of neutrinos is defined as $(m_{\nu_{d}})_i=(y_{1n})_{ii} v / \sqrt{2} $ while $M_i = M_{ii} \delta_{ij}$ is the Majorana mass matrix. The definitions of the Cabibbo-Kobayashi-Maskawa (CKM) and Pontecorvo-Maki-Nakagawa-Sakata (PMNS) matrices are \begin{eqnarray}
U_{CKM}&=& U^\dagger_{u_L} U_{d_L} ~,~\,\\
U_{PMNS} &=& U^\dagger_{e_L} U_{\nu_L}~.~\,	
\end{eqnarray}

The couplings $y_{f\phi}$ are defined as
\small{\begin{eqnarray}\label{yf}
(y_{fh_2})_{ij} &=& \frac{(m_f)_i}{v}	 c_{11} \delta_{ij}+ \frac{(y_{2f})_{ij}}{\sqrt{2}}c_{21} ~,~\,\nonumber\\
(y_{fh})_{ij} &=& \frac{(m_f)_i}{v}	 c_{12} \delta_{ij}+ \frac{(y_{2f})_{ij}}{\sqrt{2}}c_{22} ~,~\,\nonumber\\
(y_{fh_1})_{ij} &=& \frac{(m_f)_i}{v}	 c_{13} \delta_{ij}+ \frac{(y_{2f})_{ij}}{\sqrt{2}}c_{23} ~,~\,\nonumber\\
(y_{fs_1})_{ij} &=& i  \frac{(y_{2f})_{ij}}{\sqrt{2}} \cos{\alpha}~,~\,\nonumber\\ (y_{fs_2})_{ij} &=& i  \frac{(y_{2f})_{ij}}{\sqrt{2}} \sin{\alpha}~.~\,
\end{eqnarray} }

The couplings $y_{n\phi}$ of active-sterile neutrino states with the scalars are defined as
\small{\begin{eqnarray} \label{nhcoupling}
(y_{nh_2})_{ij} &=& \frac{(m_{\nu_D})_i}{v}	 c_{11} \delta_{ij}+ \frac{(y_{2n})_{ij}}{\sqrt{2}}c_{21} ~,~\,\nonumber\\
(y_{nh})_{ij} &=& \frac{(m_{\nu_D})_i}{v}	 c_{12} \delta_{ij}+ \frac{(y_{2n})_{ij}}{\sqrt{2}}c_{22} ~,~\,\nonumber\\
(y_{nh_1})_{ij} &=& \frac{(m_{\nu_D})_i}{v}	 c_{13} \delta_{ij}+ \frac{(y_{2n})_{ij}}{\sqrt{2}}c_{23} ~,~\,\nonumber\\
(y_{ns_1})_{ij} &=& i  \frac{(y_{2n})_{ij}}{\sqrt{2}} \cos{\alpha}~,~\,\nonumber\\ (y_{ns_2})_{ij} &=& i  \frac{(y_{2n})_{ij}}{\sqrt{2}} \sin{\alpha}~.~\,
\end{eqnarray}}

And the couplings between two sterile neutrinos and the scalars, $y_{nn\phi}$ are defined as
\small{\begin{eqnarray}
(y_{nnh_2})_{ij} &=&  \frac{(y_{sn})_{ij}}{\sqrt{2}}c_{31} ~,~\,\nonumber\\
(y_{nnh})_{ij} &=&  \frac{(y_{sn})_{ij}}{\sqrt{2}}c_{32} ~,~\,\nonumber\\
(y_{nnh_1})_{ij} &=&  \frac{(y_{sn})_{ij}}{\sqrt{2}}c_{33} ~,~\,\nonumber\\
(y_{nns_1})_{ij} &=& -i  \frac{(y_{sn})_{ij}}{\sqrt{2}} \sin{\alpha} ~,~\,\nonumber\\ (y_{nns_2})_{ij} &=& i  \frac{(y_{sn})_{ij}}{\sqrt{2}} \cos{\alpha}~.~\,
\end{eqnarray}}

So far, we have presented the general framework of the model without assuming any particular parameter space in mind. In the next three sections, Secs.~\ref{sec:neutrino mass}-\ref{sec:light scalar}, we want to generate a particular parameter space relevant for the rest of the paper.


 \section{neutrino masses and mixings} \label{sec:neutrino mass}
We study the mixings between the active and sterile neutrino states and the generation of neutrino masses in this section. The sterile neutrinos will generically mix with the active states and produce six neutrino eigenstates. The masses of the  three lightest eigenstates can be determined by the type-I seesaw mechanism~\cite{Minkowski:1977sc, Yanagida:1979as, GellMann:1980vs, Mohapatra:1979ia}. The part of the Lagrangian from the  Eq.~\ref{physicalyukawa}, which is responsible for  the masses of the  neutrinos, is given by
\small{ \begin{eqnarray} -\mathcal{L}_{\text{neutrino}} &=& (m_{\nu_{d}})_i (\bar{\nu}_{L_i}n_{R_i}+\bar{n}_{R_i}\nu_{L_i}) \nonumber\\&~&+\frac{1}{2} M_i (\bar{n}^c_{R_i}n_{R_i} + \bar{n}_{R_i} n^c_{R_i}) \nonumber\\ \nonumber\\
&=& \frac{1}{2}  \left( \bar{\nu}^C_{L_i} ~\bar{\eta}_{R_i}  \right) \left( \begin{array}{cc}
 0&  (m^T_{\nu_d})_i \\   (m_{\nu_d})_i &M_i    \end{array} \right) \left( \begin{array}{c} \nu_{L_i} \\  n^C_{R_i} \end{array}  \right) \nonumber\\&~&  +\text{H.c.} ~.~\,\end{eqnarray}}
 
 The  Dirac-Majorana mass matrix of neutrinos is given the $6 \times 6$ matrix  \begin{equation}  M^{D+M}_i= \left( \begin{array}{cc}
 0&  (m^T_{\nu_d})_i \\ \\  (m_{\nu_d})_i &M_i    \end{array} \right)  ~.~\,\end{equation}
 The mass matrix $M^{D+M}_i$ can be diagonalized by blocks~\cite{10.1143/PTP.64.2278, Schechter:1981cv}, up to corrections at the order of  $M_i^{-1} (m_{\nu_d})_i$,  under the assumption that all the eigenvalues of $M_i$ are  much larger than the eigenvalues of $(m_{\nu_d})_i$ 
 \begin{equation} \mathcal{W}^T M^{D+M}_i \mathcal{W} \simeq \left( \begin{array}{cc}
 (M_{\text{light}})_i & 0 \\ \\  0 &(M_{\text{heavy}})_i    \end{array} \right)  ~,~\,\end{equation}
where the $6 \times 6 $~diagonalizing matrix $\mathcal{W}$ is given by \begin{equation} \mathcal{W} \simeq  \left( \begin{array}{cc}
 1-\frac{1}{2}R R^{\dagger} & R^{\dagger} \\ \\  -R &  1-\frac{1}{2} R^{\dagger} R \end{array} \right)  \end{equation} with $R= M_i^{-1} (m_{\nu_d})_i$. The $3\times3$ light and heavy neutrino mass matrices are given by 
 \begin{eqnarray} m_{\nu_i} &=& (M_{\text{light}})_i = -(m^T_{\nu_d})_i M_i^{-1} (m_{\nu_d})_i ~,~\,\nonumber\\ m_{n_i} &=& (M_{\text{heavy}})_i = M_i  ~.~\,\end{eqnarray}\\
 We redefine $\nu_i$ and $n_i$ as the physical light active neutrinos and heavy  sterile neutrinos, respectively. The masses $m_{\nu_i}$ are not known experimentally because the neutrino oscillations are only sensitive to the differences, $m_{\nu_i}^2-m_{\nu_j}^2$. In normal hierarchy scenario, $i.e.$, assuming  $m_{\nu_1} \ll m_{\nu_2} < m_{\nu_3}$, the two mass square differences determined from the oscillation data~\cite{deSalas:2017kay} is given by $\Delta m_{21}^2 = (7.05 - 8.24) \times 10^{-5}$~eV$^2$ and $\Delta m_{31}^2 = (2.334 - 2.524) \times 10^{-3}$~eV$^2$. Therefore, there are  at least two non-zero $m_{\nu_i}$. Assuming the lightest neutrino to be massless, we get $m_{\nu_i} \simeq (0, 8.66 \times 10^{-3}, 0.05)$~eV. In Table~\ref{table:neutrino mass }, we show two  typical BPs that can generate the tiny $m_{\nu_i}$,  $m_{n_{2,3}} \sim \mathcal{O}(100)$~MeV range, and $m_{n_1} \sim \mathcal{O}(10)$~keV. Another important quantity is the mixing angle between the active-sterile states. The mixing parameters can be defined as $\theta_{ij} = M_i^{-1} (m_{\nu_d})_i (U^\dagger_{nR})_{ij}$. We also define  $\theta^2 \equiv \sum_{ij} |\theta_{ij}|^2$, and estimate it for the two BPs in Table.~\ref{table:neutrino mass }. A more detail treatment of low scale type-I seesaw can be found in Ref.~\cite{Branco:2019avf}.

 \begin{table}[h]
 \captionsetup{justification   = RaggedRight,
             labelfont = bf}
 \caption{ \label{table:neutrino mass } The parameters of two typical BPs  which are needed to generate 3 light and 3 heavy neutrinos in the normal hierarchy scenario. }
\begin{ruledtabular}
\begin{tabular}{ llll }
BP & \begin{tabular}{@{}c@{}} $M_i$ \\ (MeV) \end{tabular}  & \begin{tabular}{@{}c@{}} $(m_{\nu_d})_i$ \\ (GeV) \end{tabular}  & $\theta^2$ \\[5 pt] \hline
&&&\\
BP1 &$(0.002,420,10)$ & $(0, 1.9 \times 10^{-6}, 1.58 \times 10^{-4})$  & $6 \times 10^{-9}$\\
&&&\\
BP2 & $(0.007,380, 640)$ & $(0, 1.81 \times 10^{-6}, 5.62 \times 10^{-6})$   & $10^{-11}$ \\
 \end{tabular}
\end{ruledtabular}
\end{table}

  
 \section{Dark matter} \label{sec:dark matter}

The lightest candidate  of the heavy sterile neutrinos $n_1$ can be the DM candidate in this model if we take $m_{n_1} \simeq \mathcal{O}(1-10)$~keV. These particles can be produced at high temperature  in the early Universe but never in thermal equilibrium due to their very weak interaction strength.  These massive neutral particles are not protected by any symmetry from decaying into the lighter SM states but can have a lifetime longer than the age of the Universe controlled by the  active-sterile mixing parameter. The decay of sterile neutrinos puts bounds on the mixing parameter.  The dominant decay channel of $n_1$ would be $n_1 \rightarrow 3\nu$ through active-sterile neutrino mixing and weak interaction of $\nu$. Another possible decay channel for the given  mass range could be $n_1 \rightarrow \nu(h_1^* \rightarrow \gamma \gamma)$, where the $h_1$ decays to $2\gamma$ final state through a muon loop. But the choice of $(m_{\nu_d})_1 = 0$ forbids the channel as $(y_{nh_1})_{11} $ is directly proportional to $(m_{\nu_d})_1$. The decay width of $n_1$ decaying into $3\nu$ is given by~\cite{Pal:1981rm, Barger:1995ty}\\
  \begin{eqnarray} \Gamma_{n_1} &=& \frac{G_F^2 m^5_{n_1} \theta^2}{96 \pi^3} \nonumber\\ &\simeq& \frac{\theta^2}{1.5 \times 10^{14}~\text{sec}} \left( \frac{m_{n_1}}{10 ~\text{keV}}\right)^5 ~.~\,\end{eqnarray}

The lifetime of $n_1$ is defined as $\tau_{n_1} = 1/ \Gamma_{n_1}$. The decay of $n_1$ into $3\nu$ final state is not protected by any symmetry, therefore, to contemplate $n_1$ as a DM candidate, we need to make sure that it is long-lived enough. To make it long-lived we require $ \tau_{n_1} \gg ~t_U $, where $t_{U} = 4.4 \times 10^{17}$~sec~\cite{Aghanim:2018eyx} is the age of the Universe. This gives a bound on $\theta^2$~\cite{} as follows \begin{equation} \theta^2 \ll 3.4 \times 10^{-4}   \left( \frac{10 ~\text{keV}}{m_{n_1}} \right)^5 ~.~\,\end{equation}

The sterile neutrinos are neutral under the SM gauge symmetry, and thus do not interact  with the other particles with known forces. Because of this reason, they were not in equilibrium in the early Universe. However, they somehow must interact with other particles to be produced in the early Universe  to be a DM candidate. Therefore, the production mechanism of $n_1$ would be model dependent. In the following, we consider two benchmark mass values of $n_1$ and discuss their production mechanism. 
\begin{enumerate}

    \item $m_{n_1} = 2$~keV :  If the mass of $n_1$ is $2$~keV, it can be produced by the non-resonant Dodelson-Widrow mechanism~\cite{Dodelson:1993je}. In this scenario, the sterile neutrinos mix with the active neutrinos and  produced at high temperatures through the mixing angle suppressed weak interactions. In the type-I seesaw scenario considered in Sec.~\ref{sec:neutrino mass}, this mixing arises generically and we estimated the mixing parameter to be $\theta^2 \simeq  6 \times 10^{-9}$ for the $2$~keV $n_1$. If we consider $n_1$ as the sole DM candidate then for a given thermal history of the Universe, the DM density is uniquely determined by $m_{n_1}$ and $\theta^2$ as follows~\cite{Kusenko:2009up} \begin{equation} \label{dodelson-widrow} \Omega_{n_1}{h}^2 \sim 0.1\left( \frac{\theta^2}{3 \times 10^{-9}} \right)\left( \frac{m_{n_1}}{3 \text{keV}} \right)^{1.8}  ~,~\, \end{equation}  where $h= .72 \pm 0.08$~\cite{Aghanim:2018eyx}. From Eq.~\ref{dodelson-widrow}, we get that for $m_{n_1} = 2$~keV,  $\theta^2$, which is needed to get the correct DM abundance, is equal to $6 \times 10^{-9}$. The peak production happens at  $T \sim 200$~MeV. This benchmark point is also favored by structure formation bounds and X-ray searches~\cite{Boyarsky:2018tvu}. 
    
    \item $m_{n_1}= 7$~keV : For $n_1$ having mass $7$~keV, we estimate $\theta^2 \simeq   10^{-11}$ by taking $M_i = (0.01, 420, 500)$~MeV . This satisfies the bounds from the X-ray search~\cite{Boyarsky:2018tvu}. For such a low mixing parameter, the $n_1$ production requires an enhancement. The Shi-Fuller resonant production mechanism~\cite{Shi:1998km} can be applied to generate $n_1$. Here, lepton asymmetry produces large enhancement due to the Mikheyev-Smirnov-Wolfenstein (MSW) eﬀect~\cite{Mikheev:1986gs, Wolfenstein:1977ue}. The DM density is determined by the lepton asymmetry and $m_{n_1}$ by~\cite{Shi:1998km} \begin{equation} \label{shi-fuller} \Omega_{n_1}{h}^2 \sim 0.1 \left( \frac{m_{n_1}}{1 \text{keV}} \right) \left( \frac{\Delta L}{0.02} \right)  ~,~\, \end{equation} where $\Delta L$ is the lepton asymmetry. To get the correct relic density for $7$~keV $n_1$, we need  $\Delta L \sim 3 \times 10^{-3}$. The lepton asymmetry can be introduced in our model by assuming CP-violation in the lepton sector. The lepton asymmetry for two scalar doublet model has been studied in Ref.~\cite{Atwood:2005bf}. The decay of $7$~keV $n_1$ can be interpreted as the source of the recently observed $3.5$~keV line in the X-ray spectra of the galaxies~\cite{Bulbul:2014sua, Boyarsky:2014ska, Boyarsky:2014jta} with $\theta^2 \simeq   10^{-11}$~\cite{Boyarsky:2018tvu}.  
\end{enumerate}  

For simplicity, we assume only real Yukawa couplings and $m_{n_1}\sim \mathcal{O}(1-10)$~keV for the rest of our analysis. The complex Yukawa couplings give us more freedom on the choice of the $(\theta^2, m_{n_1})$ parameter space.


\section{light scalar} \label{sec:light scalar}

In this section, we generate a physical scalar spectrum that has interesting phenomenological aspects. Specifically, there exists a light physical scalar with mass $\mathcal{O}(100-200)$~MeV, which interacts with the physical SM fermions through tree-level FCNCs. The rest of the physical scalar masses are chosen in  a way to avoid the LHC constraints. The values of the parameters in Eq.~\ref{potential} that serve our purpose are summarized in Table.~\ref{table:scalar params }. We also present one specific BP. We see that the scalar masses $O(100)$ GeV and couplings $\lambda_i\sim 0.01-0.1$ can give rise to the 
lightest physical scalar mass $\sim 100$ MeV.

\begin{table}[h]
\captionsetup{justification   = RaggedRight,
             labelfont = bf}
\caption{ \label{table:scalar params } The descriptions of the parameters defined in Eq.~\ref{potential}. We choose the given range of values to generate a light scalar and other heavy scalars consistent with the LHC bounds. We show one specific BP. The value of $v$ is $246$~GeV.}

\begin{ruledtabular}
\begin{tabular}{ lll }

Parameters  & \begin{tabular}{@{}c@{}} Descriptions \\ and Values \end{tabular} &  \begin{tabular}{@{}c@{}} Benchmark \\  Values \end{tabular} \\\hline

\begin{tabular}{@{}c@{}}$m_1^2, m_2^2 , m_{12}^2$ \\ $m_S^2, m_{S^\prime}^2$ \end{tabular} & \begin{tabular}{@{}c@{}}  $\sim [\mathcal{O}(100)~\text{GeV}]^2$ ,\\ $m_1^2 <0, m_{12}^2<0$ \end{tabular} & \begin{tabular}{@{}c@{}c@{}c@{}c@{}} $m_1^2 = -(88.7)^2$~GeV$^2$ \\ $m_2^2 = (497)^2$~GeV$^2$  \\ $m_{12}^2 = -(55)^2$~GeV$^2$ \\ $m_S^2 = (277.7)^2$~GeV$^2$ \\  $m_{S^\prime}^2 = (199.8)^2$~GeV$^2$\end{tabular}\\\hline

\begin{tabular}{@{}c@{}} $m_{1S}, m_{2S}$ \\ $m_{12S}$ \end{tabular} & \begin{tabular}{@{}c@{}} $\sim \mathcal{O}(100)$~GeV, \\ $m_{1S}=0, m_{12S}>0$ \end{tabular}  &  \begin{tabular}{@{}c@{}c@{}} $m_{1S}=0$ \\ $m_{2S}= 50$~GeV  \\ $m_{12S}= 50$~GeV \end{tabular}  \\\hline

 \begin{tabular}{@{}c@{}c@{}}$\lambda_1,\lambda_2 ,\lambda_3 , \lambda_4 $ \\ $\lambda_5,\lambda_6, \lambda_7 $ \\ $\lambda_S, \lambda_{12S }$ \end{tabular} & \begin{tabular}{@{}c@{}} $ \sim \mathcal{O}(0.1)$, \\ $\lambda_1, \lambda_5,\lambda_6 >0 $  \end{tabular}& \begin{tabular}{@{}c@{}c@{}} $\lambda_1 = 0.26 $\\  $ \lambda_2,\lambda_3, \lambda_4,\lambda_5, \lambda_6, \lambda_7 $\\ $ \lambda_S, \lambda_{12S } = 0.1$  \end{tabular} \\\hline
 
 $\lambda_{1S}, \lambda_{2S}$ &  $\sim \mathcal{O}(0.01)$ & $\lambda_{1S} = \lambda_{2S} = 0.01$ \\
\end{tabular}

\end{ruledtabular} \end{table}

We summarize the result of the numerical calculations of the mass spectrum in Table.~\ref{table:physical scalars},~ along with the possible final states in the detectors. Details are given in the Appendix.~\ref{numerical scalar}. One important decay channel to note is the invisible SM Higgs decay, $h \rightarrow h_1 h_1$, where $h_1$ mostly decays into $n_1 \bar{n}_1$ pairs. Lack of signals from the searches at the LHC for the invisibly decaying Higgs boson put a bound on the branching fractions, Br$(h \rightarrow \text{invisible}) < 0.24$ at $95 \%$ Confidence Level (C.L.)~\cite{Khachatryan:2016whc, Aaboud:2019rtt}. For the given parameters we find the $h h_1 h_1$ coupling to be $0.42$ and $\text{Br}(h \rightarrow \text{invisible}) = 0.01$.

\begin{table}[h]
\captionsetup{justification   = RaggedRight,
             labelfont = bf}
\caption{Brief descriptions of the physical scalar spectrum needed for our analysis. We show the values of the physical masses for the BP defined in Table~\ref{table:scalar params } as well as the mass range. }
\label{table:physical scalars}
\begin{ruledtabular}
\begin{tabular}{lll}

Particles	&	\begin{tabular}{@{}c@{}c@{}} Mass values \\ for the benchmark \\ of Table.~\ref{table:scalar params } \end{tabular}  &	 \begin{tabular}{@{}c@{}}	Possible final \\  states \end{tabular} \\\hline

\begin{tabular}{@{}c@{}c@{}} Charged scalars : \\ $h^\pm $ \\ $m_{h^\pm} \sim \mathcal{O}(500)$~GeV  \end{tabular} & $m_{h^\pm} = 500$~GeV & \begin{tabular}{@{}c@{}} $h^+ \rightarrow \bar{d}_i u_j,$ \\ $e_i^+ +$MET \end{tabular} \\\hline

\begin{tabular}{@{}c@{}c@{}c@{}} Neutral scalars :  \\ $h, h_1, h_2$ \\ $m_{h_1} \sim \mathcal{O}(.1)$~GeV \\ $m_{h_2} \sim \mathcal{O}(500)$~GeV \end{tabular} & \begin{tabular}{@{}c@{}c@{}} $m_h=125.5$~GeV, \\ $m_{h_1}=0.15$~GeV \\ $m_{h_2}=500$~GeV \end{tabular}  &  \begin{tabular}{@{}c@{}c@{}c@{}} $h, h_2 \rightarrow \bar{f}_i f_j $, \\ $\gamma \gamma, h_1 h_1$ \\ $h_1 \rightarrow e^+ e^-,$  \\ $ \bar{n}_1 n_1$ \end{tabular}  \\\hline

 \begin{tabular}{@{}c@{}c@{}c@{}} Neutral \\  pseudoscalars :  \\ $s_1, s_2$  \\ $m_{s_i} \sim \mathcal{O}(500)$~GeV \end{tabular} & \begin{tabular}{@{}c@{}} $ m_{s_1}=500$~GeV, \\ $m_{s_2}=400 $~GeV  \end{tabular}& \begin{tabular}{@{}c@{}}  $s_{1,2} \rightarrow \bar{e}_i e_j$, \\  $\bar{d}_i d_j$ \end{tabular} \\
 \end{tabular}
\end{ruledtabular}
\end{table}

For the rest of the work, the light scalar $h_1$ is taken to be lighter than the muon and it promptly decays mainly to $\bar{n}_1 n_1$ or $e^+e^-$ pair with decay widths given as
\begin{eqnarray}  \Gamma(h_1 \rightarrow \bar{n}_1 n_1) &=&  (y_{nnh_1})_{11}^2 \times\frac{m_{h_1}}{16\pi} \left( 1-\frac{4m_{n_1}^2}{m_{h_1}^2}\right)^{3/2} ~,~\,\nonumber\\ \\
  \Gamma(h_1 \rightarrow e^+e^-) &=&  (y_{eh_1})_{11}^2 \times\frac{m_{h_1}}{8\pi} \left( 1-\frac{4m_{e}^2}{m_{h_1}^2}\right)^{3/2}~.~\,\end{eqnarray}

  The total decay width of $h_1$ is  $\Gamma_{h_1} = \Gamma(h_1 \rightarrow \bar{n}_1 n_1)+ \Gamma(h_1 \rightarrow e^+e^-)$,
  and the lifetime of  $h_1$ is $\tau_{h_1}= 1/\Gamma_{h_1} $. For rest of the calculations, we choose $(y_{nnh_1})_{11} = 7 \times 10^{-5}$ and $(y_{eh_1})_{11} = 10^{-5}$. Therefore, for $m_{n_1} = \mathcal{O}(1-10)$ keV and $m_{h_1}$ in the range $100-200$ MeV, we get the lifetime of $h_1$,  $\tau_{h_1} \simeq 7 \times 10^{-14}$ sec. We also obtain \begin{eqnarray}  \text{Br}(h_1 \rightarrow \bar{n}_1 n_1 ) &\simeq& 0.95 ~,~\,\nonumber\\ \text{Br}(h_1 \rightarrow e^+e^- ) &\simeq& 0.05 ~.~\,\end{eqnarray}
  
  The different constraints relevant for a light scalar of mass $\mathcal{O}(100)$~MeV are: 
  \begin{enumerate}
  
   \item  Fixed target/ Beam dump experiment: In such experiments, $h_1$ can be produced by $e$-bremsstrahlung and subsequently decays to $\bar{n}_1 n_1$ or $e^+e^-$ pair when $m_{h_1} < 2 m_\mu$. NA64~\cite{Gninenko:2019qiv} is sensitive to the invisible final states while E137~\cite{Dobrich:2015jyk, Dolan:2017osp, Bjorken:1988as, Batell:2016ove} and Orsay~\cite{Batell:2016ove} are sensitive to $e^+e^-$ final states. In electron beam dump experiments $h_1$ can also be produced via the effective coupling  $h_1 F^{\mu \nu} F_{\mu \nu}$ through a muon loop. These experiments can constrain the parameter space in $(m_{h_1},(y_{eh_1})_{11} )$ and $(m_{h_1},(y_{eh_1})_{22} )$  planes. We show these bounds in Fig.~\ref{fig:ge} and Fig.~\ref{fig:gmu}, respectively. We also show the projections from  future experiments. This parameter space is relevant for the explanations of anomalous magnetic moments of muon and electron.
 
  \item  Kaon decay: RareKaon decay into pion and electron-positron pair/invisible states can be generated via $h_1$ because of the tree-level flavor violating quark coupling, {\it i.e.}, nonzero $(y_{dh_1})_{21}$. The process $K_L \rightarrow \pi^0 n_1 \bar{n}_1$ can mimic the $K_L \rightarrow \pi^0 \nu \bar{\nu}$ decay.  NA62~\cite{Ruggiero} and E949~\cite{Artamonov:2009sz} experiments  put bounds on $((y_{dh_1})_{21}, m_{h_1})$ parameter space. We show the bounds in Fig.~\ref{fig:koto}.  This parameter space is relevant for the explanation of the anomalous KOTO events. LSND~\cite{Aguilar:2001ty} can also put constraints on this parameter space~\cite{Foroughi-Abari:2020gju}.

  \item  B-meson decay : Rare B decays $B \rightarrow K \mu^+ \mu^-$ can occur via $h_1$ due to the tree-level flavor violation in the quark sector and can put bound from LHCb experiment~\cite{Aaij:2015tna}. Without affecting any other results of our analysis, we simply choose the coupling that generates this decay to be $(y_{dh_1})_{32}\sim0$. And then this decay is   highly suppressed through the Yukawa interactions of $h_1$ channel, and we  neglect the bounds  on the $(m_{h_1},(y_{dh_1})_{32})$ parameter space.
  
  \item  Supernova cooling, $\Delta N_{eff}$, BBN: For the mass range $m_{h_1}\sim (100-200)$~MeV, the astrophysical and the cosmological bounds are very weak~\cite{Batell:2017kty, Harnik:2012ni} and therefore we do not show them here.
    
  \item  Future experiments: We also show the projected bounds from a few future/ongoing experiments such as  FASER~\cite{Feng:2017uoz, Feng:2017vli, Batell:2017kty}, SHiP~\cite{Alekhin:2015byh, Batell:2017kty}, Fermilab $\mu$-beam fixed target~\cite{Chen:2017awl, Batell:2017kty} and NA64${\mu,e}$~\cite{Gninenko:2019qiv, Chen:2017awl}.
    
    \end{enumerate}
    
We will show the constraints in later sections as required.


\section{The Muon and Electron Anomalous Magnetic Moments}	\label{sec:g-2}
The anomalous magnetic moment of the muon, $a_\mu = (g_\mu-2)/2$ has been one of the long-standing deviations between the experimental data and theoretical predictions of the SM.
The $3.7 \sigma$ discrepancy between the experimental value~\cite{Bennett:2006fi, Tanabashi:2018oca} and theoretical prediction~\cite{Davier:2017zfy, Blum:2018mom, Keshavarzi:2018mgv, Davier:2019can}  was found to be
\begin{equation}
\label{muon} \Delta a_\mu = a_\mu^{exp}-a_\mu^{th} = (2.74\pm.73)\times 10^{-9}~.~\,
\end{equation} 
Several theoretical efforts are underway to improve the precision of the SM predictions~\cite{Aubin:2019usy, Blum:2015you, Lehner:2019wvv, Davies:2019efs, Borsanyi:2020mff} by computing the hadronic light-by-light contribution with all errors under control by using lattice QCD. Recently first such result~\cite{Blum:2019ugy} was obtained and found to be consistent with the previous predictions, indicating  a new physics explanation of the discrepancy. From the experimental side, the ongoing experiment at Fermilab~\cite{Grange:2015fou, Fienberg:2019ddu} and one planned at J-PARC~\cite{Saito:2012zz} are aiming to reduce the uncertainty. 

Recently, this has been compounded with a $2.4 \sigma$ discrepancy between the experimental~\cite{Hanneke:2010au, Hanneke:2008tm} and theoretical~\cite{Aoyama:2017uqe} values of the electron  magnetic moment $a_e$
\begin{equation} \label{electron} 
\Delta a_e = a_e^{exp}-a_e^{th} = (-8.7\pm3.6)\times 10^{-13}~.~\,
\end{equation}
This $2.4 \sigma$ discrepancy came recently from the high precision measurement of the fine structure constant, $\alpha$ using the cesium atoms~\cite{Parker:2018vye}.  Note, the deviations are in the opposite directions and $\Delta a_e / \Delta a_\mu $ does not follow the lepton mass scaling, $m_e^2/m_\mu^2 \sim 2.25 \times 10^{-5}$. A new physics solution is needed to explain them simultaneously. A few possible solutions in other contexts have been considered in literature~\cite{Davoudiasl:2018fbb, Crivellin:2018qmi, Liu:2018xkx, Dutta:2018fge, Han:2018znu, Crivellin:2019mvj, Endo:2019bcj, Abdullah:2019ofw, Hiller:2019mou, Haba:2020gkr, Kawamura:2020qxo, Bigaran:2020jil, Jana:2020pxx, Calibbi:2020emz, Chen:2020jvl, Yang:2020bmh, Hati:2020fzp}.  

\begin{figure}[h]
\centering
\includegraphics[height=5cm,width=8.3cm]{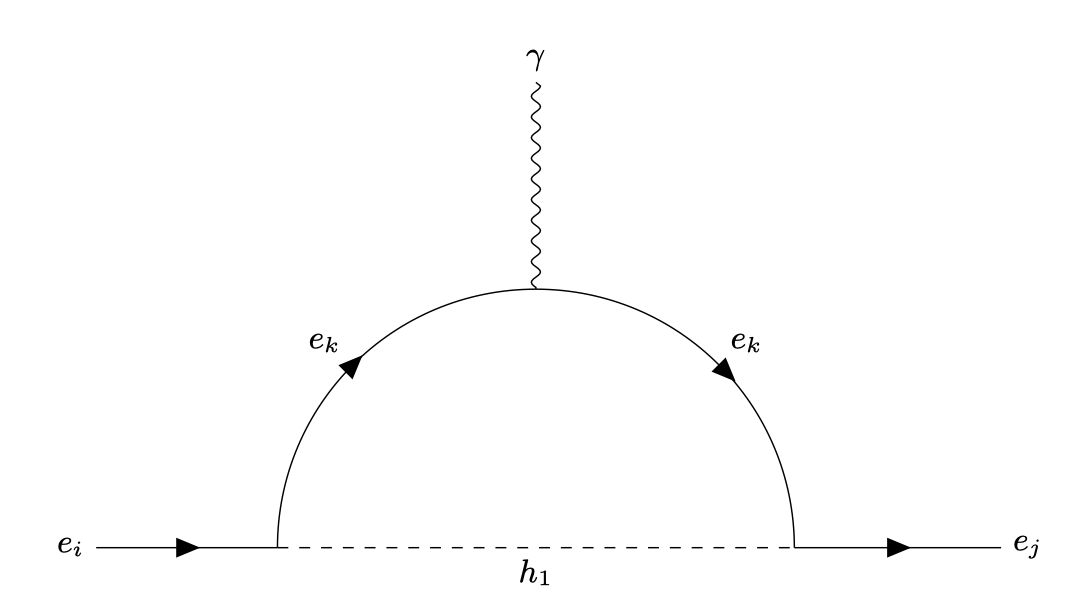}
\captionsetup{justification   = RaggedRight,
             labelfont = bf}
\caption{\label{fig:eiej} We denote an expression as {$e_i e_j, e_k$} where $e_i$, $e_j$ are the leptons in the outer legs and $e_k$ runs inside the loop.  Similar diagrams with heavier scalars are also possible which are further suppressed by the large masses. }
\end{figure}

We utilize the tree-level lepton flavor violating couplings of the light scalar $h_1$ given by Eq.~\ref{physicalyukawa} to address the issue. These couplings allow one-loop diagrams as shown in Fig.~\ref{fig:eiej} mediated by $h_1$ with different leptons inside the loop.  In general, there would be 6 different realizations of  each process with three leptons inside the loop and different chirality of  $e_i$ and $e_j$. Assuming an asymmetric Yukawa matrix,  $(y_{eh_1})_{ij}$,  we get that $\bar{e}_{iL} e_{jR} h_1$ and $\bar{e}_{iR} e_{jL}h_1$ couplings are different. We use this fact to get the opposite sign for $\Delta a_\mu$ and $\Delta a_e$. For simplicity, we further assume that some of the elements of $(y_{eh_1})_{ij}$ are zero, given in Eq.~\ref{leptonyukawa}. 

For $a_\mu$ calculation, the diagrams with muon inside the loop will  dominate. The contribution of such diagrams to the muon anomalous magnetic moments is~\cite{Leveille:1977rc}  \begin{equation}  \Delta a_{\mu \mu,\mu} =\left(y_{eh_1}\right)_{22}^2~ \frac{m_\mu^2}{4\pi^2}~ \int_0^1 dx \frac{2x^2-x^3}{x^2 m_\mu^2  +(1-x) m_{h_1}^2}~.~\,\end{equation}

\begin{figure}[h]
\centering
\includegraphics[height=5.4cm,width=8.3cm]{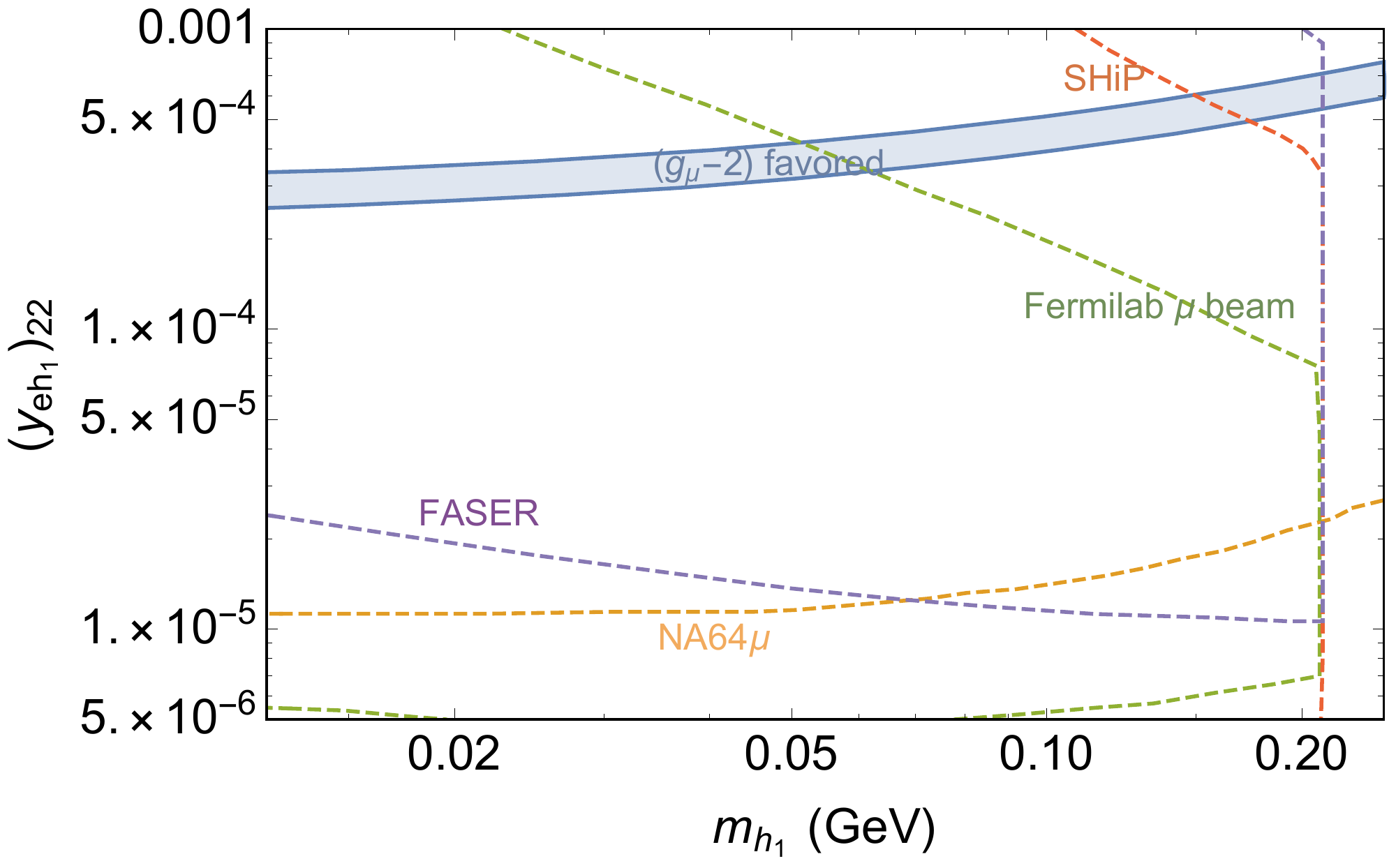}
\captionsetup{justification   = RaggedRight,
             labelfont = bf}
\caption{\label{fig:gmu} The blue shaded region shows the allowed parameter space favored by $\Delta a_\mu$.  This region of the parameter space is allowed by all muon experiments. The dotted lines show the future bounds. }
\end{figure}

In Fig.~\ref{fig:gmu}, we show the allowed parameter space in the $(m_{h_1}, (y_{eh_1})_{22})$ plane for $\Delta a_{\mu \mu,\mu} = \Delta a_\mu $. We also show relevant future bounds. This parameter space is allowed by all the muon experiment because $m_{h_1} < 2 m_\mu$.

For the electron magnetic moment both  tau and electron-induced loop diagrams are non-vanishing. The contributions to the electron anomalous magnetic moment with tau and electron inside the loop  respectively are ~\cite{Leveille:1977rc}  \begin{eqnarray} \label{aeet} \Delta a_{ee,\tau} &=& \left(y_{eh_1}\right)_{13} \left(y_{eh_1}\right)_{31}~ \frac{m_e^2}{4\pi^2}~ \nonumber\\ &~&\times~\int_0^1 dx \frac{x^2-x^3+\frac{m_\tau}{m_e}x^2}{x^2 m_e^2 +x(m_\tau^2-m_e^2) +(1-x) m_{h_1}^2} ~,~\, \end{eqnarray}
 \begin{equation} \label{aeee} \Delta a_{ee,e} =\left(y_{eh_1}\right)_{11}^2~ \frac{m_e^2}{4\pi^2}~ \int_0^1 dx \frac{2x^2-x^3}{x^2 m_e^2  +(1-x) m_{h_1}^2}~.~\,\end{equation}
 
\begin{figure}[h]
\centering
\includegraphics[height=5.4cm,width=8.3cm]{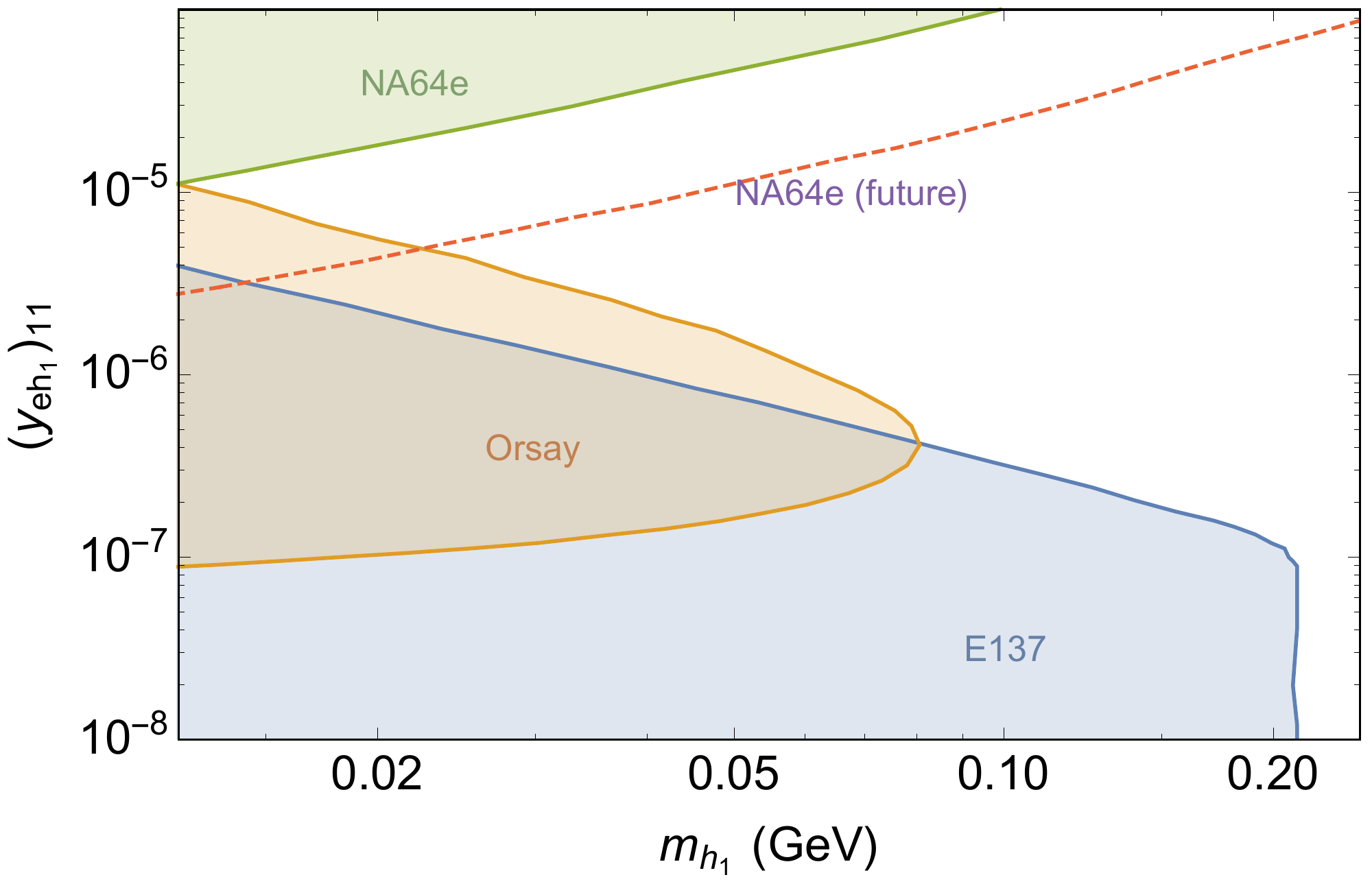}
\captionsetup{justification   = RaggedRight,
             labelfont = bf}
\caption{\label{fig:ge} The shaded regions are the excluded regions and the dotted lines show the future bounds. The value of $(y_{eh_1})_{11}) = 10^{-5}$ chosen in Sec.~\ref{sec:light scalar} falls in the allowed region for $m_{h_1} = \mathcal{O}(100-200)$~MeV.}
\end{figure}

 Note that $\Delta a_{ee,e}$ always gives positive contributions while $\Delta a_{ee,\tau}$ can be negative if one of the couplings is negative.  To explain the electron anomalous magnetic moment, we require that $\Delta a_{ee,\tau}$ gives the dominating contribution, and $\Delta a_{ee,\tau}+\Delta a_{ee,e}$ explains the deviation. In Fig.~\ref{fig:ge}, we present various constraints mentioned in Sec.~\ref{sec:light scalar} in the $(m_{h_1}, (y_{eh_1})_{11})$ plane. 
 The values of $(y_{eh_1})_{13}$ and $(y_{eh_1})_{31}$ that gives, $\Delta a_{ee,\tau} \simeq \Delta a_e$ are shown in Eq.~\ref{leptonyukawa}.
 
We choose one benchmark point which gives correct values and signs for both $\Delta a_{\mu}$ and $\Delta a_e$. 
The light scalar mass is $m_{h_1} = 140$ MeV, and  the elements of the Yukawa matrix $(y_{eh_1})_{ij}$ is given by
 \small{\begin{equation} \label{leptonyukawa} (y_{eh_1})_{ij} \simeq \left( \begin{array}{ccc}
 10^{-5}&  0& -6.8\times 10^{-4} \\  0 &5.13\times 10^{-4}& 10^{-7} \\ 3.5\times 10^{-4}&  0&  0    \end{array} \right) ~.~\,\end{equation}}
 In particular, these values do not vary much for the mass range $m_{h_1} = \mathcal{O}(100-200)$~MeV.

The Yukawa matrix in Eq.~\ref{leptonyukawa} introduces flavor violating decays mediating through the light scalar $h_1$: $\mu \rightarrow e \gamma$ with $\tau $ inside the loop,~ $\tau \rightarrow e \gamma$ with $e$~inside the loop and~$\tau \rightarrow \mu \gamma$ with $\mu$ inside the loop. The analytical expression of the branching fractions of these  decays is given in Eq.~\ref{lfvBR}. We show the values of these branching ratios using Eq.~\ref{leptonyukawa} and $m_{h_1}=140$ MeV and the corresponding experimental bounds~\cite{TheMEG:2016wtm, Aubert:2009ag} in Table \ref{table:lfv}. We find that the branching ratios are smaller than the experimental bounds.  The values do not change significantly over the mass range $m_{h_1} = \mathcal{O}(100-200)$~MeV.  

 \begin{table}[h]
 \captionsetup{justification   = RaggedRight,
             labelfont = bf}
 \caption{ \label{table:lfv} We summarize the values of different lepton flavor violating processes for the Yukawa matrix of Eq.~\ref{leptonyukawa}. We also show corresponding  experimental bounds.  }
\begin{ruledtabular}
\begin{tabular}{ llll }

Descriptions  & \begin{tabular}{@{}c@{}} Values for \\  $m_{h_1}=140$~MeV \end{tabular} & \begin{tabular}{@{}c@{}} Experimental \\  bounds \end{tabular}  \\\hline
&&\\
$\text{Br}(\mu \rightarrow e \gamma)$ & $5.75 \times 10^{-14}$ & $< 4.2 \times 10^{-13}$\\
$\text{Br}(\tau \rightarrow e \gamma)$ & $1.15 \times 10^{-11}$ & $ < 1.1 \times 10^{-7}$ \\
$\text{Br}(\tau \rightarrow \mu \gamma)$ & $1.92 \times 10^{-15}$& $< 4.5 \times 10^{-8}$\\

\end{tabular}
\end{ruledtabular}
\end{table}


\section{koto anomaly} \label{sec:koto}

The flavor changing processes like rare K meson decays, $K_L^0 \rightarrow \pi^0 \nu \bar{\nu}$ and $K^+ \rightarrow \pi^+ \nu \bar{\nu}$,  are among  the most sensitive probe for new physics beyond the SM~\cite{Buras:2015qea, Tanimoto:2016yfy, PhysRevD.96.015023, Bordone:2017lsy, Endo:2017ums, He:2018uey, Chen:2018ytc, Fajfer:2018bfj}. These decays are loop suppressed in the SM~\cite{PhysRevD.39.3322, Cirigliano:2011ny}. Any observation of such a signal would require new physics for an explanation. The SM predictions are~\cite{Buras:2015qea} \begin{eqnarray} \label{neutralKSM}
\text{Br}(K_L^0 \rightarrow \pi^0 \nu \bar{\nu})_{\text{SM}} &=& (3.00 \pm 0.30) \times 10^{-11}  \\ \label{chargedKSM}
\text{Br}(K^+ \rightarrow \pi^+ \nu \bar{\nu})_{\text{SM}} &=& (9.11 \pm 0.72) \times 10^{-11}
\end{eqnarray}
    
The KOTO experiment~\cite{ComfortKOTO, Yamanaka:2012yma} at J-PARC~\cite{Nagamiya:2012tma} and NA62 experiment~\cite{NA62:2017rwk} at CERN are dedicated to probing these processes. Recently, four candidate events were observed in the signal region of $K_L^0 \rightarrow \pi^0 \nu \bar{\nu}$ search at KOTO experiment, whereas the SM prediction is only $0.10 \pm 0.02$~\cite{Shinohara, Lin}. Out of four events, one can be suspected as a background coming from the SM upstream activity, while the other three can be considered as signals as they are not consistent with the currently known background. Given, single event sensitivity as $6.9 \times 10^{-10}$~\cite{Shinohara, Lin}, three events are consistent with \begin{equation} \label{koto19}
    \text{Br}(K_{L}^0 \rightarrow  \pi^0 \nu \bar{\nu} )_{\text{_{\scriptsize{KOTO19}}}}< 2.1^{+2.0(4.1)}_{-1.1(-1.7)} \times 10^{-9}
\end{equation} at 68(90)$\%$ C.L., including statistical uncertainties. The result includes the interpretation of photons and invisible final states as $\nu \bar{\nu}$. Note, the central value is almost two orders of magnitude larger than the SM prediction. This new result is in agreement with their previous bounds~\cite{Ahn:2018mvc} \begin{equation}
    \text{Br}(K_{L}^0 \rightarrow  \pi^0 \nu \bar{\nu} )_{\text{_{\scriptsize{KOTO18}}}}< 3.0 \times 10^{-9}~.~\,
\end{equation} 

On the other hand, the charged kaon decay searches did not see any excess events. The recent update from NA62 puts a bound~\cite{Ruggiero} \begin{equation}
    \text{Br}(K^+ \rightarrow  \pi^+ \nu \bar{\nu} )_{\text{_{\scriptsize{NA62}}}}< 2.44 \times 10^{-10}
\end{equation} at 95$\%$ C.L.,  which is consistent with the SM prediction of Eq.~\ref{chargedKSM}.   

In general, the neutral and charged kaon decays satisfy
the following Grossman-Nir (GN) bound~\cite{Grossman:1997sk}
\begin{equation}
{\rm Br}\left(K^0_L \rightarrow \pi^0 \nu {\bar \nu}\right)  \leq 4.3 \times {\rm Br}\left(K^+ \rightarrow \pi^+ \nu {\bar \nu}\right) ~,~\,
\end{equation}
which depends on the isospin symmetry and kaon lifetimes. 
The GN bound might give a strong constraint on the explanations for the KOTO anomaly. Thus, the new physics explanation for the KOTO anomaly is required to generate three anomalous events and satisfy the GN bound. Several such solutions have been proposed in the literature~\cite{Kitahara:2019lws, Fabbrichesi:2019bmo, Egana-Ugrinovic:2019wzj, Dev:2019hho, Li:2019fhz, Jho:2020jsa, Liu:2020qgx, Liao:2020boe, Cline:2020mdt, Gori:2020xvq,He:2020jzn, He:2020jly, Datta:2020auq, Foroughi-Abari:2020gju, Altmannshofer:2020pjb}.

In this work, we rely on the tree-level flavor violating couplings of the light scalar $h_1$ in the quark sector of Eq.~\ref{physicalyukawa} and invisible decay channel of $h_1$ to interpret Eq.~\ref{koto19}. The non-zero value of $(y_{dh_1})_{21}$ leads to the tree-level $s \rightarrow d$ transition through $h_1$. Thus, the neutral kaon can decay into a neutral pion and a $h_1$ through the tree-level coupling. The same coupling would allow the charged kaon to decay into a charged pion and a $h_1$. The produced $h_1$  promptly decays into either a DM pair $n_1\bar{n}_1$ or an electron pair. The decay channel $\text{Br} (K_{L}^0 \rightarrow \pi^0 n_1 \bar{n}_1)$ will mimic the $\text{Br}(K_{L}^0 \rightarrow  \pi^0 \nu \bar{\nu}) $ search signals and can account for the required branching fractions of Eq.~\ref{koto19}. 
Note that the $\text{Br}(K^+ \rightarrow \pi^+ + \text{invisible})$ bound is generally stronger except in the mass range $ \sim m_\pi \pm 25$~MeV~\cite{Ruggiero, Artamonov:2009sz, CortinaGil:2018fkc, Fuyuto:2014cya}, therefore, we choose the mass parameter $m_{h_1}$ in that range to evade the GN bound,

The non-zero coupling $(y_{dh_1})_{21}$ also gives the tree-level $K^0-\bar{K}^0$ mixing mediated via $h_1$. The contribution of this mixing to the $K_L -K_S$ mass difference can be calculated as follows
\small{\begin{eqnarray} \Delta m_K = -\frac{2 (y_{dh_1})_{21}^2}{m_{h_1}^2}\frac{f_K^2m_K^2}{12m_K}\left[ 1-\frac{m_K^2}{(m_s+m_d)^2} \right]~,~\,
\end{eqnarray}} with $\Delta m_K^{\text{exp}} = 3.52 \times 10^{-15} $~GeV~\cite{Tanabashi:2018oca}. Here. $f_K \simeq 1.23 m_\pi$ is the kaon decay constant~\cite{Tanabashi:2018oca}. For $m_{h_1} = \mathcal{O}(100-200)$~MeV, one only needs $(y_{dh_1})_{21} < 10^{-8}$ to avoid this constraint, which is obviously satisfied in the following discussions.

 The decay width of $K_{L}^0$ decaying into a neutral pion and an on-shell $h_1$ is
\small{ \begin{eqnarray} \Gamma(K_{L}^0 \rightarrow \pi^0 h_1) &=& \frac{[{\rm Re}(y_{dh_1})_{21}]^2}{16 \pi m_{K_L^0}} ~\left( \frac{m_{K^0}^2-m_{\pi^0}^2}{m_s-m_d} \right)^2 f^2(m_{h_1}^2) \nonumber\\ &~& \times  ~  \lambda^{1/2}\left(1, \frac{m_{\pi^0}^2}{m_{K_L^0}^2}, \frac{m_{h_1}^2}{m_{K_L^0}^2}\right) ~,~\, \end{eqnarray}} where $\lambda(x, y, z)=x^2+y^2+z^2-2xy-2yz-2zx$ is the triangle function, and 
the function $f(q^2)$ for the vector form factor is defined as~\cite{McWilliams:1980cz} \small{\begin{eqnarray} f(q^2)&=& f_+(0)\left( 1+\frac{\lambda_0}{m_\pi^2} q^2\right)  \end{eqnarray}}
with $f_+(0) = 0.97$ and $\lambda_0 = 1.8 \times 10^{-2}$.
 
 And the decay width of $K_{L}^+$ decaying into a charged pion and an on-shell $h_1$ is
 \begin{eqnarray} \Gamma(K^\pm \rightarrow \pi^\pm h_1) &=& \frac{|(y_{dh_1})_{21}|^2}{16 \pi m_{K^\pm}} ~\left( \frac{m_{K^\pm}^2-m_{\pi^\pm}^2}{m_s-m_d} \right)^2 f^2(m_{h_1}^2) \nonumber\\ &~& \times  ~  \lambda^{1/2}\left(1, \frac{m_{\pi^\pm}^2}{m_{K^\pm}^2}, \frac{m_{h_1}^2}{m_{K^\pm}^2}\right) ~.~\,\end{eqnarray}
 
  The  $h_1$ produced in the decay of the kaon is short-lived with typical lifetime $\tau_{h_1} \simeq 10^{-13}$ sec for the choice of the parameters in Sec.~\ref{sec:light scalar}. Now taking the energy of the produced $h_1$ to be $E_{h_1} \simeq 1.5$ GeV, we estimate the path it travels before it decays as, $\gamma c \tau_{h_1} \simeq 10^{-4} $ m. The length of the KOTO detector is $3$ m, hence $h_1$ decays inside the detector. It can promptly decay into $n_1 \bar{n}_1$ or $e^+e^-$ pair with branching fractions of $0.95$ and $0.05$, respectively. So we get
  \small{ \begin{eqnarray} \text{Br} (K_{L}^0 \rightarrow \pi^0 n_1 \bar{n}_1) &=& \frac{ \Gamma(K_{L}^0 \rightarrow \pi^0 h_1)\times \text{Br}(h_1 \rightarrow n_1 \bar{n}_1)}{\Gamma_{K_L^0}},\nonumber\\ 
  \text{Br} (K_{L}^0 \rightarrow \pi^0 e^+e^-) &=& \frac{ \Gamma(K_{L}^0 \rightarrow \pi^0 h_1)\times \text{Br}(h_1 \rightarrow e^+e^-)}{\Gamma_{K_L^0}},\nonumber \\  \end{eqnarray}}
  where $\Gamma_{K_L^0} = \Gamma_{K_L^0}^{\text{SM}}+\Gamma(K_{L}^0 \rightarrow \pi^0 n_1 \bar{n}_1)+\Gamma(K_{L}^0 \rightarrow \pi^0 e^+e^-)$ with $\Gamma_{K_L^0}^{\text{SM}} = (1.29 \pm 0.01)\times 10^{-17}$~GeV. We get similar expressions for the $K^\pm$ decays.
  
\begin{figure}[h]
\centering
\includegraphics[height=5.7cm,width=8.6cm]{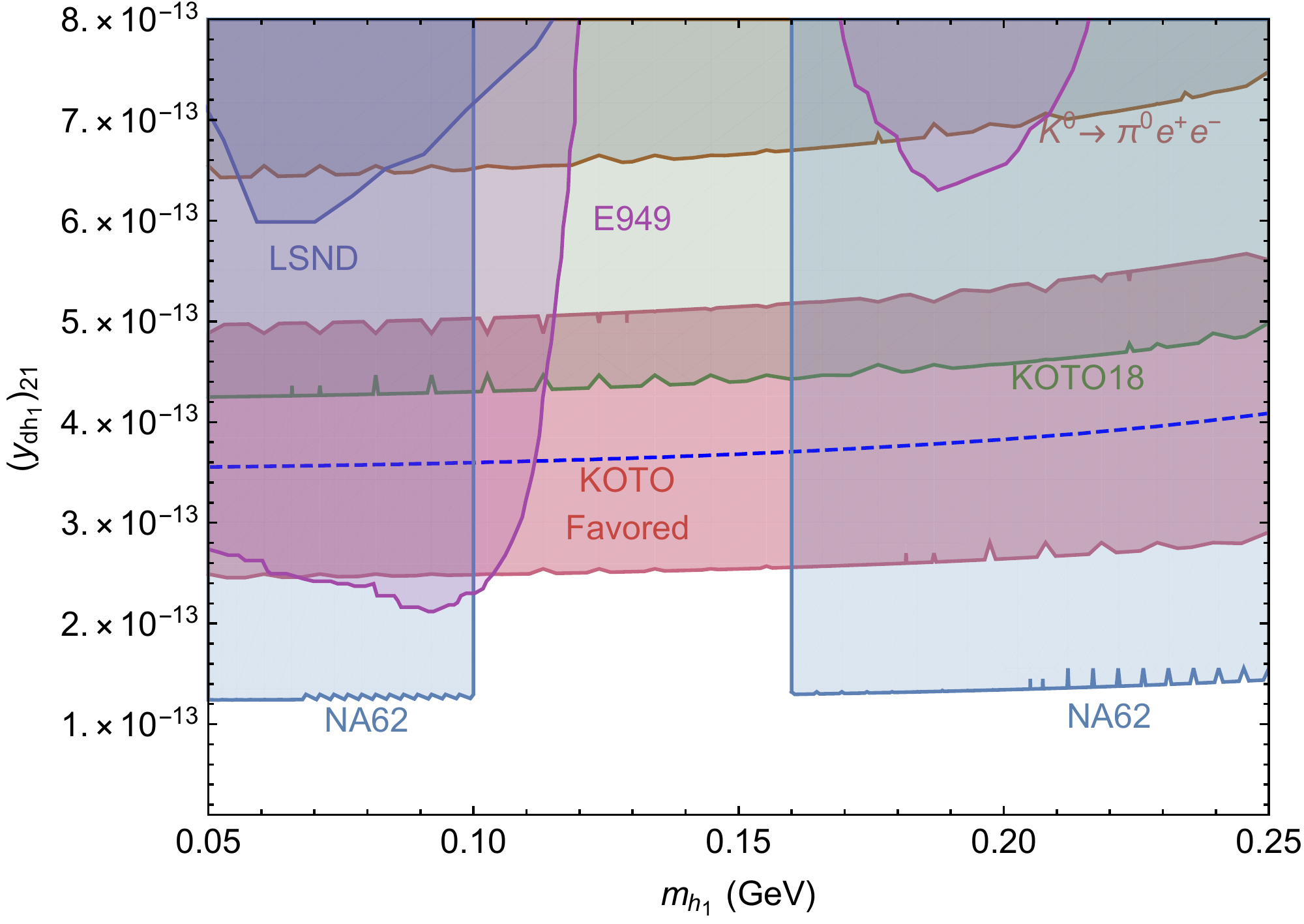}
\captionsetup{justification   = RaggedRight,
             labelfont = bf}
\caption{\label{fig:koto} \small{The pink shaded region is the parameter space favored by the KOTO anomaly in our model. The blue dashed line is the contour corresponding to the central value of the KOTO anomaly. The green contour corresponds to the KOTO18 excluded region. Contour line corresponding to the $K_L^0 \rightarrow \pi^0 e^+e^-$ decay is shown in brown. We also show the excluded region by NA62, E949 and LSND.} }
\end{figure}

  In Fig.~\ref{fig:koto}, we show the favored parameter space in $(m_{h_1}, (y_{d h_1})_{21})$ plane corresponding to the branching fraction of Eq.~\ref{koto19}. We also show the region excluded by KOTO 2018 result and $K_{L}^0 \rightarrow \pi^0 e^+e^-$ decay channel. As mentioned earlier, the KOTO favored region is allowed by the NA62 experiment, thus avoiding the GN bound. 


\section{Miniboone excess}  \label{sec:miniboone}

MiniBooNE is a Cherenkov detector consists of a $12.2$~m diameter sphere filled with 818 tonnes of pure mineral oil (CH$_2$), located at the Booster Neutrino Beam (BNB) line at Fermilab~\cite{AguilarArevalo:2008qa}. The experiment gets the neutrinos and antineutrinos flux from BNB~\cite{AguilarArevalo:2008yp}. Recently, in 2018, after taking data for 15 years, they have reported a $4.7 \sigma$ excess of $\nu_e +\bar{\nu}_e$ like events over the estimated background in the energy range $200 < E_\nu^{QE} < 1250$~MeV ~\cite{Aguilar-Arevalo:2018gpe}. The amount of combined excess events is $460.5 \pm 99.0$ corresponding to $12.84 \times 10^{20}$ protons on target in neutrino mode and $11.27 \times 10^{20}$ protons on target in antineutrino mode. This result is in tension with the two-neutrino oscillation within the standard three neutrino scenario. More recently this result was updated by MiniBooNE with  $638\pm132.8$ electron-like events (4.8$\sigma$) as the reported number of excess events corresponding to $18.75 \times 10^{20}$ protons on target in neutrino mode and $11.27 \times 10^{20}$ protons on target in antineutrino mode~\cite{Aguilar-Arevalo:2020nvw}. 

Recently, several attempts have been put forth to explain this anomaly within the context of  dark neutrino mass models using heavy sterile neutrinos and  dark gauge bosons~\cite{Bertuzzo:2018ftf, Bertuzzo:2018itn, Ballett:2018ynz, Ballett:2019cqp, Ballett:2019pyw, Abdallah:2020biq} and dark sector models with dark scalars~\cite{Datta:2020auq}. They all considered the scenario where the light neutrinos upscatter to a  heavy neutrino after coherent scattering off the nucleus and  subsequent decay of the heavy neutrino into a pair of electrons. The MiniBooNE detector cannot distinguish the electron pair. One can get the reconstructed neutrino energy using the energy and angular distribution of the mediator coming from the sterile neutrino decay~\cite{Martini:2012fa}. Recently, it was shown that parameter space needed for the explanation of MiniBooNE data in the dark gauge boson models are constrained by CHARM-II data~\cite{Arguelles:2018mtc}, because the scattering cross-section get enhanced for large neutrino energy. The scalar mediator models have the advantages as for similar parameters, as the scattering cross-section is much smaller~\cite{Datta:2020auq}.

\begin{figure}[h]
\centering
\includegraphics[height=5.3cm,width=6.7cm]{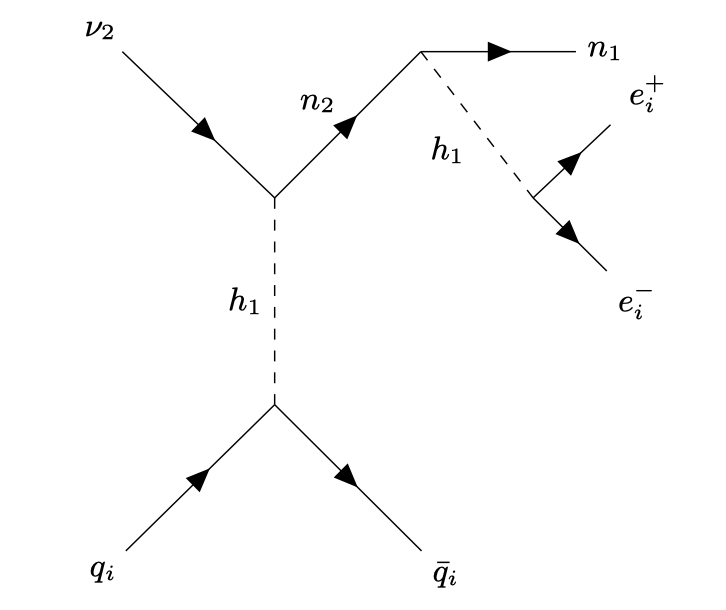}
\captionsetup{justification   = RaggedRight,
             labelfont = bf}
\caption{\label{fig:mini} The Feynman diagram  for the upscattering process $\nu A \rightarrow n A$ that contributes to the cross-section that generates the MiniBooNE excess events in our model. }
\end{figure}

In the framework of our model, the heavy sterile neutrino $n_2$ can be produced from the upscattering process: $\nu_2 A \rightarrow n_2 A$ mediated through the light scalar $h_1$ as shown Fig.~\ref{fig:mini}. The $\nu_2 A$ scattering being coherent is enhanced by $\sim A^2$. The produced $n_2$  promptly decays into $n_1$ and an on-shell $h_1$, which subsequently decays into a pair of $e^+e^-$ with $\text{Br}(h_1)\rightarrow e^+e^- \simeq 5 \%$. Taking the typical energies, $E_{n_2}, E_{h_1} \sim 1$~GeV, we  estimate the length of the path they travel before decay as $l_{n_2} \le 10^{-4}$~m and $l_{h_1} \simeq 10^{-4}$~m.  

As both the heavy neutrino $n_2$ and the light scalar $h_1$ decay promptly, we can write the  total number of events observed due to this process as
\begin{eqnarray} N_{\text{event}} &=& f_{\text{exp}} \int_{E_{\nu_{\text{min}}}}^{E_{\nu_{\text{max}}}} dE_\nu \Phi(E_\nu) \int_{E_{R_{\text{min}}}}^{E_{R_{\text{max}}}} dE_R \nonumber\\ \nonumber\\ &~&\times ~ \frac{d\sigma(E_R, E_\nu)}{dE_R} \times \text{Br}(h_1 \rightarrow e^+e^-) ~,~\,\end{eqnarray}
where $f_{\text{exp}}$ is a  factor which involves the numbers of protons on target, exposure, effective area of the detector and depends on the  experiments; $E_R$ is the nuclear recoil energy; $E_\nu$ is the incoming neutrino energy; and $\Phi(E_\nu)$ is the incoming neutrino  flux from  the BNB. Therefore, $f_{\text{model}} = N_{\text{event}} / f_{\text{exp}}$ is the model-dependent part.

The differential scattering cross-section of $\nu A \rightarrow n A$ is given by
\begin{eqnarray} \frac{d\sigma}{dE_R} &=&[Z f_p +(A-Z) f_n]^2 \frac{(y_{nh_1})_{22}^2}{16 \pi E_{\nu}^2} \nonumber\\ \nonumber\\&\times& \frac{(m_{n_2}^2+2 m_A E_R)(2 m_A + E_R)}{(m_{h_1}^2 + 2 m_A E_R)^2} F^2(E_R)  ~,~\,\end{eqnarray}
where $m_A$ is the mass of the target nucleus; $Z$ and $A-Z$ are the proton and neutron numbers of the target nucleus; $F(E_R)$ is the nuclear form factor~\cite{Helm:1956zz, Engel:1991wq}; and the factors $f_{p,n}$ are defined as~\cite{Falk:1999mq} \small{\begin{equation} \frac{f_{p,n}}{m_N}=\sum_{q=u,d,s}f_{T_q}^{(p,n)}\frac{f_q}{m_q}+\frac{2}{27}\left(1-\sum_{q=u,d,s}f_{T_q}^{(p,n)}\right) \sum_{q=c,b,t}\frac{f_q}{m_q}. \end{equation}} We take, $f_{(u,d)} = (y_{(u,d) h_1})_{11}$ and $f_{s,c,b,t} = 0$. The constants $f_{T_u}^{(p)}$, $f_{T_d}^{(p)},f_{T_u}^{(n)}$ and $ f_{T_d}^{(n)}$  are taken to have the values 0.020, 0.041, 0.0189, and 0.0451, respectively~\cite{Alarcon:2011zs, Alarcon:2012nr, Crivellin:2013ipa, Hoferichter:2015dsa, Junnarkar:2013ac}. 

\begin{figure}[h]
\centering
\includegraphics[height=5.3cm,width=8.4cm]{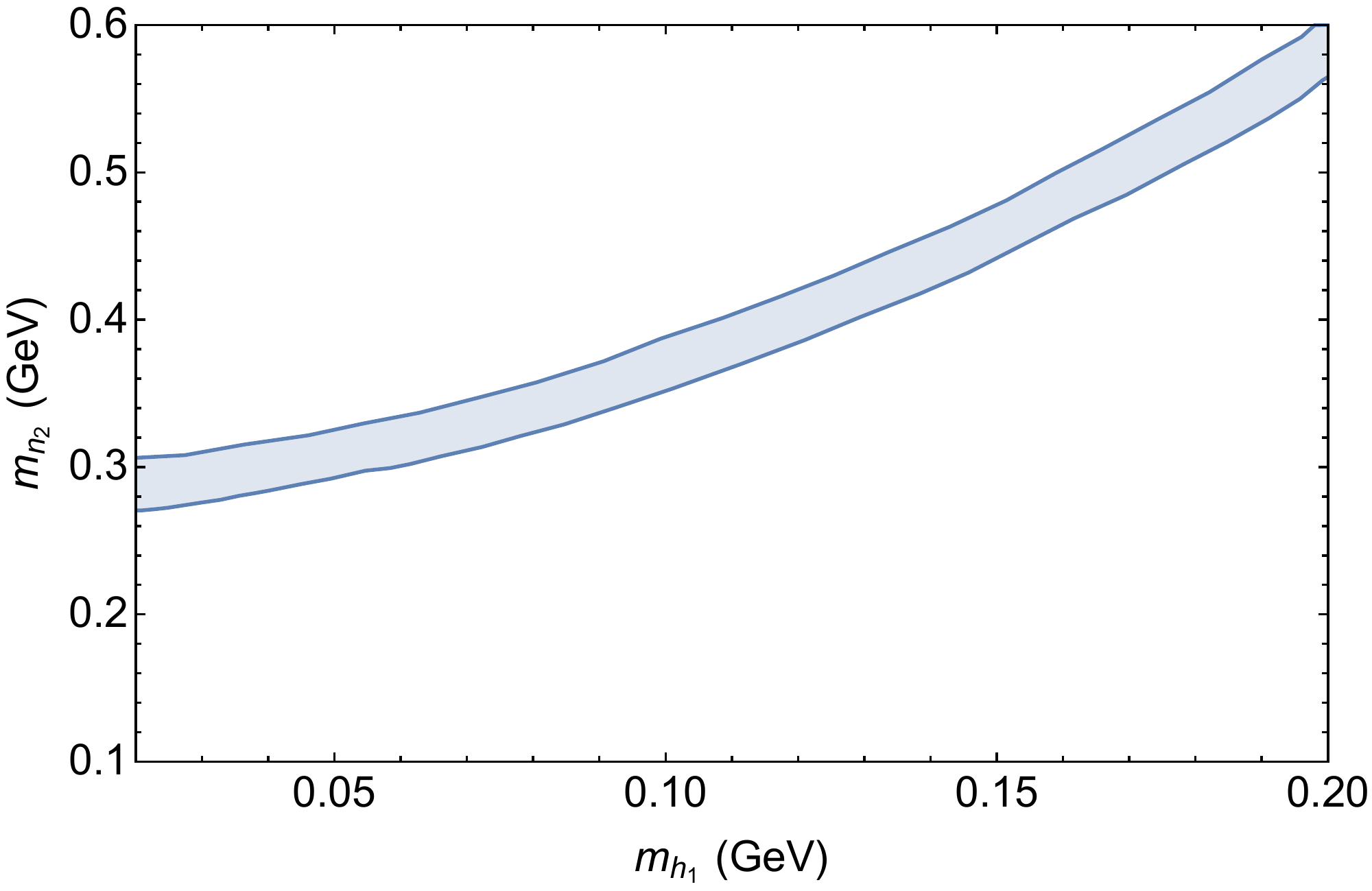}
\captionsetup{justification   = RaggedRight,
             labelfont = bf}
\caption{\label{fig:miniboone} The shaded region is the allowed parameter space in the $(m_{h_1}, m_{n_2})$ plane which gives  the desired numbers of total events. We take the couplings: $(y_{nh_1})_{22}=6.1 \times 10^{-2}$, $(y_{uh_1})_{11}=5.0 \times 10^{-6}$ and $(y_{dh_1})_{11}=5.0 \times 10^{-6}$.}
\end{figure}

Fig.~\ref{fig:miniboone} shows the allowed values of $n_2$ masses for $m_{h_1} = \mathcal{O}(100-200)$~MeV to generate the MiniBooNE events given the couplings : $(y_{nh_1})_{22}=6.1 \times 10^{-2}$, $(y_{uh_1})_{11}=5.0 \times 10^{-6}$ and $(y_{dh_1})_{11}=5.0 \times 10^{-6}$. This is consistent with the neutrino masses and mixing in our model as shown in Table.~\ref{table:neutrino mass }.

\begin{figure}[h]
\includegraphics[height=5.3cm,width=8.4cm] {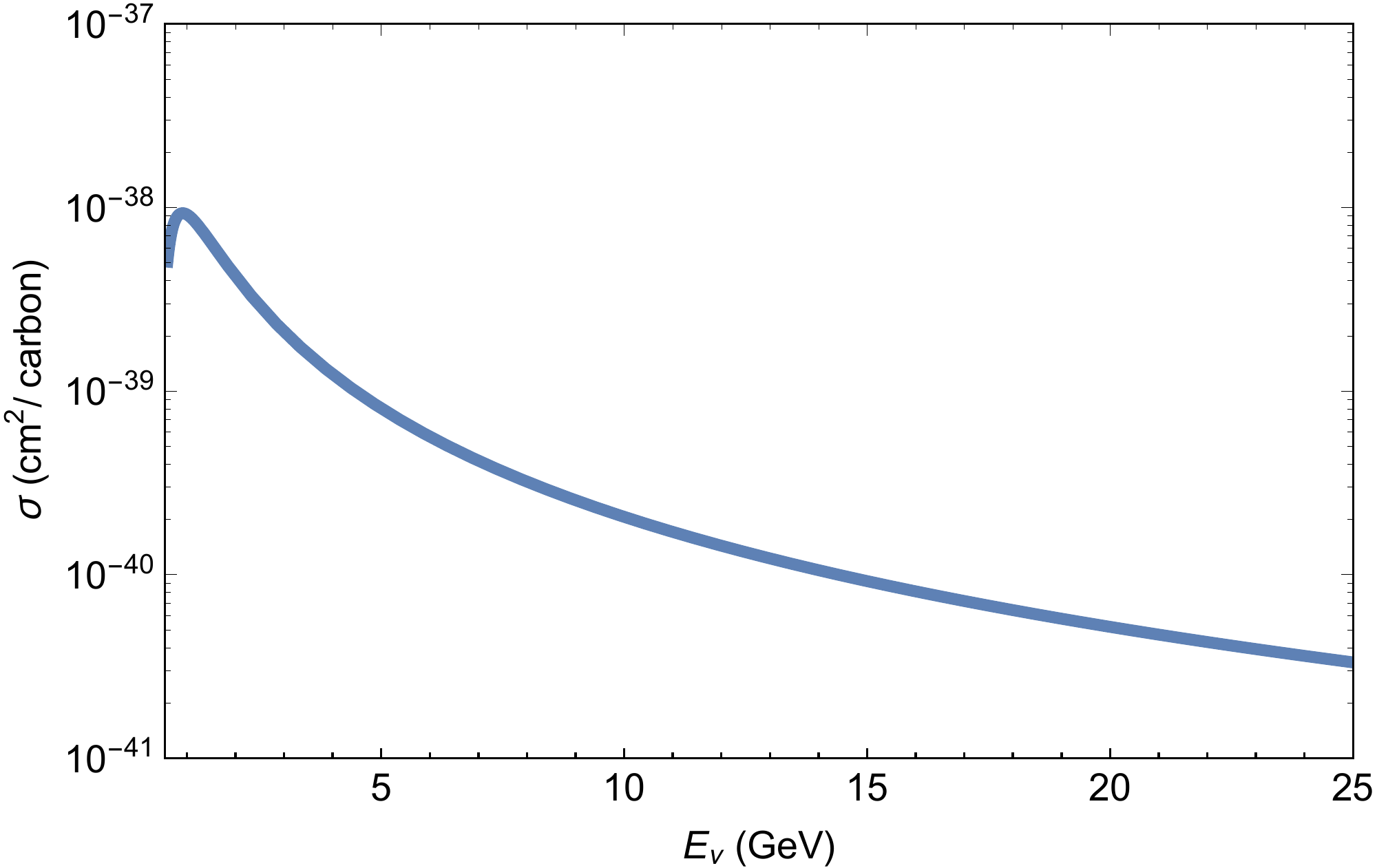}
\captionsetup{justification   = RaggedRight,
             labelfont = bf}
\caption{\label{fig:minicrosssection} The line shows the cross section as a function of the incoming neutrino energy for the BP: $m_{n_2}=420 ~\text{MeV}$, $m_{h_1}= 140 ~\text{MeV}$, $(y_{nh_1})_{22}=6.1 \times 10^{-2}$, $(y_{uh_1})_{11}=5.0 \times 10^{-6}$, and $(y_{dh_1})_{11}=5.0 \times 10^{-6}$.}
\end{figure}

 We choose one typical benchmark point $m_{n_2} = 420 ~\text{MeV and }~m_{h_1}= 140 ~\text{MeV}$ to show the scattering cross-section as a function of the incoming neutrino energy in Fig.~\ref{fig:minicrosssection}. Note, the cross-section is small at the relevant incoming neutrino energy, $E_{\nu_\mu} = 20 $~GeV~\cite{Layda:1991bm} of the CHARM-II experiment~\cite{DeWinter:1989zg, Geiregat:1992zv, Vilain:1994qy}, therefore gives no excess events~\cite{Datta:2020auq}. It was shown recently~\cite{Brdar:2020tle} that, if the decay length of the produced sterile neutrino $n_2$ in the upscattering has decay length $l_{n_2} \le 10^{-4}$ m, then the scalar mediated process does not produce any excess events in T2K ND280~\cite{Abe:2011ks, Kudenko:2008ia, Assylbekov:2011sh, Amaudruz:2012esa, Abe:2012av, Abe:2019kgx} and MINER$\nu$A~\cite{Wolcott:2016hws, Park:2015eqa, Wolcott:2015hda, Valencia:2019mkf} experiments. We also verify that our model-dependent part $f_{\text{model}}$ is consistent with other dark gauge bosons~\cite{Bertuzzo:2018itn, Arguelles:2018mtc} or dark scalar models~\cite{Datta:2020auq}. We show the estimated number of excess events for a few benchmark points in Table \ref{table:observables}.
 
\section{discussions} \label{sec:discussions}

We have considered  a general framework of the scalar singlet-doublet extension of the SM scalar sector and added three sterile neutrinos. We have generated a very interesting physical particle mass spectrum which has rich phenomenological consequences. In particular, the particles that play central role in our analysis are: one light scalar with mass $m_{h_1} \sim \mathcal{O}(100-200)$~MeV, the lightest sterile neutrino with  mass $m_{n_1} \sim \mathcal{O}(1-10)$~keV and the next-to-lightest sterile neutrino with mass $m_{n_2} \sim \mathcal{O}(400)$~MeV. The lightest sterile neutrino $n_1$ can be a viable DM candidate. $n_1$ with a mass of 7 keV can explain the 3.5 keV line in the X-ray search.  We have shown that one can get tiny neutrino mass and DM relic abundance in this model as well.

The main focus of the work was to address a few of the recent experimental puzzles: anomalous magnetic moments of both muon and electron; KOTO anomalous events and excess events found in the MiniBooNE neutrino experiment. The tree-level flavor violating couplings of the light scalar to the leptons enable us to explain the $(g-2)_{\mu,e}$ using one-loop diagrams.  And the flavor violation in the quark sector allows the Kaon to decay at tree level. All the flavor violations associated with the scalars in this model appear at the tree level. The MiniBooNE, on the other hand, requires the production of heavy sterile neutrino from the light scalar mediated neutrino-nucleus scattering. Note, the tree-level FCNC of the light scalar and the decay of the light scalar to electron-positron pair and  a pair of lightest sterile neutrinos connect all three puzzles.
 
\begin{table}[h]
\captionsetup{justification   = RaggedRight,
             labelfont = bf}
 \caption{ \label{table:benchmarks} Three BPs are shown, for which we calculate the different observables quantities, and can account for three anomalies. }
\begin{ruledtabular}
\begin{tabular}{ llll }

Parameters  & BP1 & BP2 & BP3 \\\hline
$m_{h_1}(\text{MeV})$ & $130$ & $140$& $150$\\
$m_{n_1}(\text{keV})$ & $2$ & $3$ & $2$\\
$m_{n_2}(\text{MeV})$ & $435$& $420$&$440$\\
$(y_{eh_1})_{22}$ & $5 \times 10^{-4}$ & $4.75 \times 10^{-4}$ & $5.5 \times 10^{-4}$\\$(y_{eh_1})_{13}$ & $-3.5 \times 10^{-4}$ & $-6 \times 10^{-4}$& $-6.8 \times 10^{-4}$\\$(y_{eh_1})_{31}$ & $6.8 \times 10^{-4}$& $4 \times 10^{-4}$&$3.5 \times 10^{-4}$\\$(y_{dh_1})_{21}$ & $3 \times 10^{-13}$ & $3.5 \times 10^{-13}$&$4 \times 10^{-13}$\\

\end{tabular}
\end{ruledtabular}
\end{table}

We showed that the parameter space found in Sec.~\ref{sec:neutrino mass}-\ref{sec:light scalar} can explain these anomalies simultaneously. We found that the light scalar mass is tightly constrained for the explanation of the KOTO anomaly which emerges in a large region in the allowed parameter space. We chose three BPs in the allowed region of the parameter space and summarize them in Table~\ref{table:benchmarks}.  For all these BPs, we  fix  the coupling constants: $(y_{nnh_1})_{11}=7 \times 10^{-5}$, $(y_{eh_1})_{11}= 1\times 10^{-5}$, $(y_{nh_1})_{22}=6.1 \times 10^{-2}$, $(y_{uh_1})_{11}=5.0 \times 10^{-6}$, and $(y_{dh_1})_{11}=5.0 \times 10^{-6}$. We summarize the observables in Table~\ref{table:observables}. These BPs can also explain neutrino masses and mixing angles.

\begin{table}[h]
\captionsetup{justification   = RaggedRight,
             labelfont = bf}
\caption{ \label{table:observables}  The observables corresponding to the three BPs. }
 \begin{ruledtabular}
\begin{tabular}{ llll }

Observables  & BP1 & BP2 & BP3 \\ \hline

$\Omega _{n_1} h^2$ & $0.1$ & $0.1$& $0.1$\\

$\Delta a_\mu \times 10^{-9}$ & $2.67 $ & $2.27 $ & $2.86 $\\

$\Delta a_e \times 10^{-13}$ & $-8.43 $& $-8.50$&$-8.43 $\\

$Br(K_{L}^0 \rightarrow  \pi^0 n_1 \bar{n}_1 ) \times 10^{-9}~$ &$1.42 $ &$1.91 $ &$2.47 $\\

$Br(K_{L}^0 \rightarrow  \pi^0 e^+ e^- )  \times 10^{-11}~ $ & $ 5.81 $&$ 7.82$ &$ 1.01 $\\

$N_{\text{event}}~ (\nu+\bar{\nu} ) $ & $671$ & $644$ & $497$ \\

\end{tabular}
\end{ruledtabular}
\end{table}
The parameter space associated with the explanation of MiniBooNE excess is not constrained by the existing data from MINER$\nu$A, CHARM-II and T2K ND280 data due to the scalar mediator. If however, in future, the MiniBooNE data requires the scalar mediator mass to be $\leq 100$ MeV then the KOTO explanation would be in tension with the model. In that case, we would need more than one light scalar to satisfy both KOTO and MiniBooNE anomalies. Further, since this model has three sterile neutrinos, the lightest sterile neutrino mass can be $\sim 1$ eV which satisfies the oscillation data  whereas the second to lightest neutrino ($\sim 400$ MeV) can explain the low energy excess in the MiniBooNE data.

The light scalar model we presented in this paper appears to be quite effective in explaining the DM content, neutrino masses, and various anomalies. This model would be investigated as we obtain more results on these anomalies from KOTO, $(g-2)_{\mu, e}$, MicroBooNE etc. along with various ongoing and upcoming experiments, e.g., NA64$\mu$,e; FASER, SHiP, Fermilab $\mu$-beam etc. and various lepton flavor violating rare decays.

\begin{acknowledgments}
We are grateful to Sudip Jana, Bill Louis and Yongchao Zhang for useful discussions. We thank Vedran Brdar for carefully reading our paper and helping us to debug one of the figures. B.D., and S.G. are supported in part by the DOE Grant No. DE-SC0010813. 
T.L. is supported in part by 
the Projects 11875062 and 11947302 supported by the
National Natural Science Foundation of China, and by
the Key Research Program of Frontier Science, CAS. We have used the TikZ-Feynman~\cite{Ellis:2016jkw} package to generate the Feynman diagram of Fig.~\ref{fig:eiej} and \ref{fig:mini}.
\end{acknowledgments}

\appendix

\section{Higgs Basis Transformation} \label{appendix:higgs basis}
We consider two complex scalar doublet $H_{1,2}$ and one scalar singlet $H_S$ singlet with the following quantum numbers under $SU(2)_L \times U(1)_Y$ gauge symmetry \begin{equation}
H_1\sim (2,1/2),~~~~~H_2\sim (2,1/2),~~~~~H_S\sim (1,0)~.~\,
\end{equation} The most general charge conserving vev's are
\begin{eqnarray}
\small{\braket{H_1} = \left( \begin{array}{c} 0 \\ \frac{v_1}{\sqrt{2}} \end{array}  \right),~ \braket{H_2} = \left( \begin{array}{c} 0 \\ \frac{v_2}{\sqrt{2}} \end{array}  \right),~\braket{H_S} = \frac{v_3}{\sqrt{2}}}~.~\, \end{eqnarray}

We redefine the neutral components of the Higgs fields by rotating via a Unitary matrix $U$ in such a way that only one scalar doublet will develop a non-zero vev. The neutral components of the new Higgs fields can be written as \begin{equation} \phi^0_a = \sum_b U_{ab} H^0_b ~,~\,\end{equation} where, $a,b = 1,2$ and $S$. The Unitary matrix $U$ is given as \begin{equation} U = \left( \begin{array}{ccc} \frac{v_1}{v} & \frac{v_2}{v} & \frac{v_3}{v} \\ -\frac{v_2}{v} & \frac{v_1}{v} & 0 \\ -\frac{v_3}{v} & 0 & \frac{v_1}{v} \end{array} \right) ~.~\,\end{equation} It is easy to see that the vev's of the new Higgs fields are given by \begin{equation} \braket{\phi^0_1} = \left( \begin{array}{c} 0 \\ \frac{v}{\sqrt{2}} \end{array}  \right),~ \braket{\phi^0_2} =0, ~\braket{\phi^0_S} =0 ~,~\,\end{equation} where $v = \left( v_1^2 +v_2^2 +v_3^2\right)^{1/2}$. Therefore, only one doublet will control the spontaneous electroweak gauge symmetry breaking and the generation of the SM fermion masses. 

\section{Numerical Calculation of Scalar Spectrum} \label{numerical scalar}

Some details about the numerical analysis of Sec.~\ref{sec:light scalar} is given here. Given the benchmark values of the parameters in Table~\ref{table:scalar params }, one can follow Eqs.~\ref{chargedscalar}-\ref{etadef} to calculate the mixing of the scalar interaction states and the masses of the physical scalars. The summary of the masses is given in Table~\ref{table:physical scalars}. In particular, the physical neutral scalars are given by $h_i = (U_{R}^{-1})_{ij} \rho_j$: 
\small{\begin{eqnarray} \label{numphyshiggs}
h_2 &=& 0.056~\rho_1 +0.995~ \rho_2 + 0.081~\rho_3 ~,~\,\nonumber\\ 
h &=& 0.997~\rho_1 - 0.053~\rho_2 - 0.035~\rho_3 ~,~\,\nonumber\\
h_1 &=& 0.030~\rho_1 - 0.083~\rho_2 +0.996~\rho_3 ~.~\,
\end{eqnarray}}
Eq.~\ref{numphyshiggs} tells us that the heavy scalar $h_2$ mostly comes from the second doublet $\phi_2$, while the SM Higgs is associated with the doublet $\phi_1$. The light scalar $h_1$ mostly comes from the singlet. These mixing elements also enter into Eq.~\ref{yf}. The mixing angle between the pseudoscalars are $\alpha = 5.44^\degree$ and the physical states are given by
\small{\begin{eqnarray} \label{Pseudo-numphyshiggs}
s_1 &=& 0.995~\eta_2 -0.094~ \eta_S ~,~\,\nonumber\\ 
s_2 &=& 0.094~\eta_2 +0.995~\eta_S  ~.~\,
\end{eqnarray}}
The physical scalars $s_1$ and $s_2$ are mostly associated with the doublet $\phi_2$ and $\phi_S$, respectively.

\section{Calculation of $e_i \rightarrow e_j \gamma$} \label{appendix:lfv}

The most general expression for the branching fraction of the process $e_i \rightarrow e_j \gamma$ for a light scalar mediator of Fig.~\ref{fig:eiej} is given by \begin{eqnarray} \label{lfvBR} \text{Br}(e_i \rightarrow e_j \gamma) &=& \frac{\Gamma(e_i \rightarrow e_j \gamma)}{\Gamma(e_i \rightarrow e_j \bar{\nu}_j \nu_i)}  \nonumber\\ 
&=& \frac{3 \alpha}{8 \pi G_F^2 m_{e_i}^2}\left( 1- \frac{m_{e_j}^2}{m_{e_i}^2}\right) \left[ (y_{eh_1})_{ik} (y_{eh_1})_{kj} \right]^2 \nonumber\\&~& \times~\frac{I_1(m_{e_i},m_{e_j},m_{e_k},m_{h_1})}{I_2(m_{e_j}^2/m_{e_i}^2)} ~,~\,\end{eqnarray} 
 where the lepton $e_k$ runs inside the loop. The function $I_1(m_{e_i},m_{e_j},m_{e_k},m_{h_1})$ comes from the partial decay width $\Gamma(e_i \rightarrow e_j \gamma)$ whereas $I_2(m_{e_j}^2/m_{e_i}^2)$ comes from $\Gamma(e_i \rightarrow e_j \bar{\nu}_j \nu_i)$ . The definitions of the functions $I_1$ and $I_2$ respectively are \small{ \begin{eqnarray}&~& I_1(m_{e_i},m_{e_j},m_{e_k},m_{h_1})  = \int_0^1 dz \int_0^{1-z} dy \nonumber\\  &~&~~~~~\times \frac{yz(m_{e_j}-m_{e_i})-(z-1)(zm_{e_i}+m_{e_k})}{z(y+z-1)m_{e_i}^2-yzm_{e_j}^2+(1-z)m_{e_k}^2+zm_{h_1}^2} ~,~\,\nonumber\\ 
&~&I_2\left( \frac{m_{e_j}^2}{m_{e_i}^2} \right) = 1-8\frac{m_{e_j}^2}{m_{e_i}^2}+8\frac{m_{e_j}^6}{m_{e_i}^6}-\frac{m_{e_j}^8}{m_{e_i}^8}\nonumber\\  &~&~~~~~~~~~~~~~~~~~ +12\frac{m_{e_j}^4}{m_{e_i}^4}\ln\left( \frac{m_{e_i}^2}{m_{e_j}^2} \right) ~.~\,\end{eqnarray}}

 \bibliographystyle{apsrev4-1.bst}
\bibliography{miniboone.bib}
 
\end{document}